\documentclass[aps,prd,twocolumn,superscriptaddress]{revtex4-1} 
\usepackage{fullpage, amsmath, xparse, amssymb, mathtools, graphicx, braket,xcolor,verbatim}
\usepackage{natbib}
\usepackage[colorlinks=true,linkcolor=blue,citecolor=blue]{hyperref}

\usepackage{multirow}

\begin{document}
\title{Investigating the use of field solvers for simulating classical systems}
\author{Andrew Eberhardt}
\email{Corresponding author. \\ aeberhar@stanford.edu}
\affiliation{Kavli Institute for Particle Astrophysics and Cosmology, Menlo Park, 94025, California, USA}
\affiliation{Physics Department, Stanford University, Stanford, California, USA}
\affiliation{SLAC National Accelerator Laboratory}
\author{Arka Banerjee}
\affiliation{Kavli Institute for Particle Astrophysics and Cosmology, Menlo Park, 94025, California, USA}
\affiliation{Physics Department, Stanford University, Stanford, California, USA}
\affiliation{SLAC National Accelerator Laboratory}
\author{Michael Kopp}
\affiliation{Nordita, KTH Royal Institute of Technology and Stockholm University, Roslagstullsbacken 23, SE-106 91 
Stockholm, Sweden}
\author{Tom Abel}
\affiliation{Kavli Institute for Particle Astrophysics and Cosmology, Menlo Park, 94025, California, USA}
\affiliation{Physics Department, Stanford University, Stanford, California, USA}
\affiliation{SLAC National Accelerator Laboratory}
\begin{abstract}

We explore the use of field solvers as approximations of classical Vlasov-Poisson systems. This correspondence is investigated in both electrostatic and gravitational contexts. We demonstrate the ability of field solvers to  be excellent approximations of problems with cold initial condition into the nonlinear regime. We also investigate extensions of the Schr\"odinger-Poisson system that employ multiple stacked cold streams, and the von Neumann-Poisson equation as methods that can successfully reproduce the classical evolution of warm initial conditions. We then discuss how appropriate simulation parameters need to be chosen to avoid interference terms, aliasing, and wave behavior in the field solver solutions. We present a series of criteria clarifying how parameters need to be chosen in order to effectively approximate classical solutions.   

\end{abstract}

\maketitle

\section{Introduction}
Field solvers have proved useful for simulating a number of physical phenomena. These solvers have been applied to study two distinct types of systems. The first are those systems in which the quantum nature of the constituents is relevant, such as fuzzy dark matter \cite{Shukla, Schive, MoczBECDM, Suarez, Schive:2014,Hu, Davies, Levkov, Vicens, Davies:2019, Mocz, Woo_2009}, quantum plasma \cite{Anderson}, small electronics \cite{Haas, Tan}, nonlinear optics \cite{Paredes}, and superfluids \cite{Roberts2001}, among others. The second type are systems where the field solver approximates the classical solution, such as cold dark matter \cite{Kopp, Garny, Widrow, Davies, Uhlemann, Mocz, Johnston, Tigrak_2011} and classical plasma \cite{Navy, NGUYEN1981431, SUH, bertrand_1980}. One motivation in the latter application is to map out and understand the regimes of correspondence between field solvers and traditional classical methods \cite{Haas, Kopp, Garny, Widrow, Davies, Navy, SUH, NGUYEN1981431, Uhlemann}. This correspondence is of particular interest because of field solvers' potential ability to reduce the computational resources necessary to simulate certain systems \cite{Widrow, Mocz, Haas, Kopp} as well as circumvent some of the problems associated with classical solvers \cite{Davies}. 

 The most commonly used field solver, i.e. the Schr\"odinger-Poisson (SP) method, at the most basic level, evolves an array of complex numbers in time using the Schr\"odinger-Poisson equations. This array can be thought of as representing a ``wave function" in some basis, but is more often interpreted as a complex classical field \cite{Widrow}. As with true quantum wave functions, however, one can interpret the square of the absolute value of a given element as the probability density at that position in the chosen basis, typically the position or velocity bases \cite{Kopp}. In the limit of a large number of particles, as in classical systems, this probability density corresponds to a physical density. Further, the derivative of the phase can be interpreted as a position of that density in the reciprocal space of the basis. For example, when represented in the position basis, the derivative of the phase corresponds to the velocity of the stream at that position \cite{Widrow}. This wave function is advanced in time using the Schr\"odinger equation, where the potential term appearing in the equation is computed using the Poisson equation and the density is given by the square of the absolute value of the wave function in the position basis \cite{Widrow}. Note that Fourier transforms relate position and momentum spaces, allowing the wave function, and by extension, the field solver methods, to represent both spatial density and velocity dispersion \cite{Kopp}. An approximation of a classical phase space representation can then be created with the Husimi transform \cite{Ball, Kopp, Widrow, Davies, Uhlemann, Widrow}, allowing a full description of our system. The correspondence between the Husimi distribution and coarse grained Vlasov distribution is also of mathematical interest \cite{ATHANASSOULIS}.

Investigation of the classical limit of these solvers has been a topic of active research in the literature for some time \cite{Haas, Kopp, Garny, Widrow, Davies, Navy, SUH, NGUYEN1981431, Uhlemann}. The specific way in which the solutions from the field solvers reduce to the classical limit, as well as the limitation and benefits of using these solvers in quantum scenarios is not fully understood, but it has been shown that field methods are quite successful in simulating some classical plasma problems involving cold initial conditions \cite{Widrow, Kopp, Haas, Anderson}. 

When used as approximations of classical systems, field solvers present a scenario which a $2\times n$ dimensional phase space can be represented by an $n$ dimensional array of complex numbers \cite{Widrow, Davies, Kopp}. This means that for appropriate systems, the quantum representation may use fewer computational resources. Understanding the limitations and benefits of field solvers may provide new means to investigate these systems via simulation. Further, classical methods are known to exhibit shot noise in low density regions \cite{Davies, Sellwood_2012}. Field techniques which use continuous fields can, in principle, circumvent this problem \cite{Davies}.

In this work, we investigate the field solvers' correspondence with classical solvers by looking at the results of a number of idealized test problems. We discuss both how the classical limit is achieved, as well as the cause of deviations. We demonstrate that separation of streams in phase space is necessary to approximate the classical solutions using the standard Schr\"odinger method. Importantly, we show that the limit of validity can be extended to warm initial conditions, in which streams may be poorly resolved, by using modifications to this technique. 

We organize the paper as follows. In Section \ref{sec:Physics}, we discuss the background physics of the system being simulated. Section \ref{sec:Solvers} contains a description of each solver used in the paper, including a description of how to create initial conditions, the employed algorithms, and a discussion of how field solvers correspond to classical counterparts. Section \ref{sec:TestCases} contains a series of test cases involving both gravitational and electrostatic problems in which we investigate the ability of field solvers to reproduce classical results. In this section, we start with simple cold distributions and move to initial conditions with velocity dispersion to track how and why the field solver can fail, and strategies that can improve the convergence toward a classical solution. We include a discussion of the results in Section \ref{sec:Discussion}, focusing on the limitations and behavior of the field solvers and the regimes in which they can be used as a classical approximation. Conclusions regarding the overall utility of these methods and future directions are presented in Section \ref{sec:Conclusion}.

\section{Background} \label{sec:Physics}

The fundamental systems being discussed here are collisionless plasma or self-gravitating ``dust" which source their own potential. The potential is given by the Poisson equation,
\begin{equation}
    \nabla^2 V = C \rho \, ,
\end{equation}
where $V$ is the potential and $\rho$ is a density. We will refer to $C$ as the Poisson constant. When $C<0$, the potential is an electrostatic one and the relevant density is a charge density. When $C>0$, the potential is gravitational and the relevant density is a mass density. All the simulation methods presented in this paper can be used to solve both types of problems.

A key tool in the analysis of these systems will be the investigation of the distribution function on phase space $f_t(x,u)$ which gives the density of particles at time $t$ with velocity $u$ at position $x$. We will be interested in two classes of solvers. The first class is our classical solver which solves the Vlasov equation written as
\begin{equation}
    \frac{\partial f}{\partial t} = q\nabla V \cdot \frac{\partial f}{\partial u} - u \cdot \nabla f \, .
\end{equation}
where $q=e/m$ is the charge to mass ratio in the electrostatic case and $q=1$ in the gravitational case. Qualitatively, this solver tracks and evolves the position and velocity of a large number of particles and uses their spatial distribution to calculate a density \cite{Hockney}.

\subsection{Schr\"odinger-Poisson system}

The second class of solvers will include our field solvers. The most basic and well investigated in the literature is the Schr\"odinger-Poisson solver which solves the nonlinear Schr\"odinger equation: 
\begin{equation}
    i \hbar \frac{\partial \psi}{\partial t} = -\frac{\hbar^2}{2m} \nabla^2 \psi + q m V \psi \, .
\end{equation}
Qualitatively, this solver tracks and evolves a complex field \cite{Davies, Widrow, Uhlemann}, which can be represented in position or velocity space. When represented in position space, the amplitude of the field is interpreted as the spatial density and the derivative of the phase is interpreted as the velocity of the density at that position \cite{Widrow}. The extent to which this system, and its extensions, can be used to approximate the classical solver is the main focus of this work. 

The field representation of our system makes phase space representation difficult, as in general a quantum system has no direct classical phase space analog \cite{Ball}. However, there exist phase space representations which maintain many of the properties of classical phase space and work as approximations of classical phase space. These can be useful tools to analyze the correspondence between the classical and quantum systems. 

Phase space representations of quantum states typically have properties that deviate from those of classical phase space representations. Wigner and Husimi distributions have both been used \cite{Kopp, Garny, Widrow, Davies, SUH, Uhlemann, bertrand_1980, BERTRAND1983415, Mocz}. However, Wigner distributions are only everywhere positive when the quantum state is a superposition of Gaussians and the Husimi distribution only recovers the coarse-grained moments of the given distribution \cite{Ball}. Here we use the Husimi distribution, which has the advantage of being everywhere positive, as an approximation of classical phase space. When interpreting the Husimi distribution as an approximation of the classical phase space it is important to recall that structures in phase space below the characteristic area of the smoothing kernel cannot be resolved.

The Husimi function represents a mapping from a state to phase space. The phase space itself is composed of basis states for a given position ($q$) and momentum ($p$) in phase space given as follows \cite{Ball} (in the position basis of our wave function)
	\begin{equation}
	\label{basis}
	\braket{x|q,p} \equiv(2 \pi s^2)^{-1/4}e^{-(x-q)^2/4s^2} e^{ipx/\hbar}\,,
	\end{equation}
in which $s \in \mathbb{R}$ is a free parameter representing the spatial width of the Gaussian smoothing kernel. These states are obviously not orthogonal but can be made to form a resolution to the identity as follows
	
	\begin{equation}
	(2\pi \hbar)^{-1} \int \ket{p,q}\bra{p,q} dp dq = I \,.
	\end{equation}
	
In general the phase space distribution ($f^H_t(q,p)$) can be obtained by taking the modulus squared of the wave function evaluated in the $(q,p)$ basis states. Note, however, that,
	\begin{align}
		&\left| \psi(q) \right|^2 \neq \int_{\text{All}} f_H(q,p) dp \, ,\\
		&\left| \psi(p) \right|^2 \neq \int_{\text{All}} f_H(q,p) dq \, .
	\end{align}

Instead the right-hand sides of these equations correspond to the coarse grained spatial and momentum densities with smoothing scales $s$ and $\hbar/2s$, respectively.  

Finding the probability distribution for each field solver then simply involves taking an inner product of the basis states with the wave functions or density matrices, described in Table \ref{tab:update}. For the Schr\"odinger system this is done as
\begin{widetext}
\begin{align}
    f^H_t(q,p) &= (2\pi \hbar)^{-1}\left| \int dx (2 \pi s^2)^{-1/4}e^{-(x-q)^2/4s^2} e^{ipx/\hbar} \psi(x) \right|^2 \nonumber \\
    &= (2\pi \hbar)^{-1}\left| \braket{q,p|\psi} \right|^2 \, .
\end{align}
\end{widetext}

\subsection{von Neumann-Poisson system}

It is also possible to evolve a statistical ensemble of classical fields using traditionally quantum methods. For this we turn to the density matrix formalism and use the von Neumann (VN) equation,

\begin{equation}
    \frac{\partial P}{\partial t} = \frac{-i}{\hbar} [H,P] \, , 
\end{equation}
where $P$ is the density operator representing our ensemble of fields and $H$ is the same Hamiltonian used in the Schr\"odinger-Poisson system,
\begin{equation}
    H = \frac{p^2}{2m} + mV \, .
\end{equation}
A thorough discussion of the von Neumann equation and matrix density formalism can be found in many graduate level quantum textbooks, for example \cite{Ball}. This formalism is used to describe the evolution of statistical mixtures quite frequently in other subfields in physics \cite{Berman, Verstraete, Zwolak}.
Our density operator is written as
\begin{equation}
    P = \sum_i w_i \ket{\psi_i} \bra{\psi_i} \, ,
\end{equation}
where $w_i$ is the probability weight of the $i$th field $\psi_i$. In general, a change of basis can be performed on the density operator as follows \cite{Ball}:
\begin{equation}
    \tilde{P} = \braket{f|P|f'} = \sum_i \sum_{i'} \braket{f|i}\braket{i|P|i'} \braket{i'|f'} \, , 
\end{equation}
where $\tilde{P}$ is the density operator represented in the basis new states $\ket{f}$, and $\braket{i|P|i'}$ represents the density operator represented in the original basis states $\ket{i}$. 

This extension of the Schr\"odinger-Poisson equation allows the evolution of a dispersion of different velocity streams without having them interfere with one another. We will show that this is useful for simulating warm initial conditions. In general, this formalism should be more accurate if the phase space we are trying to represent is not a one-to-one function of $q$ and $p$, as is the case with multiple spatially overlapping streams.

\section{Solvers} \label{sec:Solvers}

In this section, we discuss the different numerical methods employed in this paper, including the procedures for setting up consistent initial conditions for the different solvers, along with the update rules for the variables of each method. We also discuss the conditions under which these solvers reproduce the classical time evolution of Vlasov-Poisson systems, along with the conditions under which they deviate from the classical analog. The system of units is discussed in the Appendix \ref{sec:units}. 

All solvers use a symplectic drift-kick-drift update scheme that works as follows \cite{Springel}:

\begin{enumerate}
    \item Half step position update 
    \item Calculate potential 
    \item Full step momentum update 
    \item Half step position update
\end{enumerate}

\begin{table*}

\begin{center}

\begin{tabular}{| c | c | c | c | c |}

\hline
 & Classical & Schr\"odinger & Multiple Hilbert spaces & von Neumann \\ 
 
\hline \hline

Position & $r_{t+ \Delta t} = r_{t} + v \, \Delta t $ & $\Tilde{\psi}_{t + \Delta t} = U^T_t(\Delta t)\Tilde{\psi}_t$ & $\Tilde{\psi}_{t + \Delta t,i} = U^T_t(\Delta t)\Tilde{\psi}_{t,i}$ & $ \Tilde{P}_{t + \Delta t} = U_t^T(\Delta t) \, \Tilde{P}_t \, U_t^{T\dagger}(\Delta t) $ \\

\hline

Momentum & $v_{t+ \Delta t} = v_{t} + a \, \Delta t $ & $\psi_{t + \Delta t} = U^V_t(\Delta t)\psi_t$ & $\psi_{t + \Delta t, i} = U^V_t(\Delta t)\psi_{t,i}$ & $P_{t + \Delta t} = U^V_t(\Delta t) \, P_t \, U^{V\dagger}_t(\Delta t) $ \\

\hline
Density & CIC deposit & $|\psi|^2$ & $\sum_{i = 1}^{N_s} |\Psi_i|^2$  & $\textrm{Diag}(P)$ \\

\hline

$f_t(q,p)$ & CIC deposit & $(2\pi \hbar)^{-1}\left| \braket{q,p|\psi} \right|^2$ & $(2\pi \hbar)^{-1} \sum^{n_s}_{i} \left| \braket{q,p|\Psi_i} \right|^2$  &   $(2\pi \hbar)^{-1} \braket{q,p|P|q,p}$ \\
\hline
\end{tabular}

\caption{\label{tab:update} Shown for each solver is the full step position update, momentum update, density deposit, and phase space rules. A comparison of Schr\"odinger method and its extensions reveals the similarities between all the field solvers. At its core each of the field solvers involves update via unitary operators diagonal in either the momentum or position basis. Likewise, the phase space distribution for each is simply given as a the norm of the fields after a change of basis into the phase space eigenstates.}

\end{center}

\end{table*}

\subsection{Classical solver}
\label{sec:classical}
\subsubsection{Initial conditions}

For the classical solver, the system is discretized in terms of particles. We generate initial conditions for this solver simply by choosing an initial position and velocity for each particle, sampled from the initial distribution function. Generally we will discuss initial conditions by describing their constituent streams. Here a stream refers to a collection of particles all with the same velocity. Streams may be spatially uniform or given some spatial density dependence. Generally we will initialize streams with some spatial or velocity perturbation.

For our classical solver we first divide up the total number of particles $N_{\rm part}$ into individual streams. The streams are populated with particles according to their Boltzmann weights i.e. the number of particles in the $i$th stream is 
\begin{align}
    N_{\rm part}^i = N_{\rm part} \frac{e^{-v_i^2/v_{\rm th}^2}}{Z} \, ,
\end{align}
where $N_{\rm part}^i$ is the total number of particles in stream $i$, $v_i$ is the mean velocity of the stream,  and $v_{\rm th}$ is the thermal velocity. $Z$ is the partition function of the system defined as
\begin{align}
    Z = \sum_{i} e^{-v_i^2/v_{\rm th}^2} \, .
\end{align}
To determine the position of particles within a given stream, we compute the cumulative distribution function (CDF) of the initial density distribution. Note that if all particles have the same mass that a mass density function normalized by the total mass, $M_{\rm tot}$ , can be interpreted as a probability distribution function. Therefore the CDF can be written as follows
\begin{align}
    {\rm CDF}(r) \equiv \frac{1}{M_{\rm tot}} \int_{-\infty}^{r} \rho (x) dx \,.
\end{align}
The position for the $j$-th particle in the stream can be found by inverting this CDF, i.e. the particle is assigned position $r_j$ which satisfies
\begin{align}
    {\rm CDF}(r)|_{r=r_j} = \frac{j+1}{N_{\rm part}^i+2} \, .
\end{align}
For the plasma simulations, the initial spatial density is constant and so this procedure implies that the particles are spaced uniformly over the simulation length.
Finally, the velocity for particle $j$, starting in stream $i$ is assigned as follows
\begin{align}
    v_j = v_i + v_p \sin\bigg(2 \pi \frac {r_j}{\lambda}\bigg) \, ,
\end{align}
where $v_p$ controls the magnitude of the perturbation of velocity about the mean velocity $v_i$ of the stream and $\lambda$ its wavelength.

\subsubsection{Update}

Once the initial conditions have been set up, we have two arrays of length $N_{\rm part}$ representing the positions and velocities which need to be updated at every time step. The update rule is described in Table \ref{tab:update} and is as follows:
\begin{align} \label{cl_update}
    r_{t+ \Delta t} &= r_{t} + v \, \Delta t/2 \textrm{  (position update half step)} \nonumber \\
    &\textrm{(calculate acceleration)} \\
    v_{t+ \Delta t} &= v_{t} + a \, \Delta t \textrm{  (velocity update full step)} \nonumber \\
    r_{t+ \Delta t} &= r_{t} + v \, \Delta t/2 \textrm{  (position update half step)} \nonumber \, ,
\end{align}
which is a common leap frog integrator known to be symplectic \cite{Springel, wanner}.

The accelerations, $a$, are computed as follows: the particle positions are used to create a density field $\rho(x)$ on the grid using the particle-in-cell deposition scheme. For these simulations we use a simple top hat deposit of grid spacing width or cloud-in-cell (CIC) deposit \cite{Hockney}, we will define a top hat deposit as follows:
\begin{align}
    T_i(x) = \Theta\left(x - r_i + \frac{\Delta x_g}{2}\right) - \nonumber \\
    \Theta\left(x - r_i - \frac{\Delta x_g}{2}\right) \, ,
\end{align}
where the density is then constructed as a sum over the particles as follows
\begin{equation}
    \rho(x) = q \sum_i T_i(x) \,.
\end{equation}
Here $q$ is the particle mass or charge, depending on whether the system is interacting under gravity or electromagnetism, $r_i$ the position of the $i$th particle, and $\Delta x_g \equiv L/N$ is the width of a grid cell.

The computation of the acceleration from the density field is done in Fourier space, so we consider the Fourier transform of the density,
\begin{align}
     \rho'(k) = \int \rho(x) e^{ikx} dx \,.
\end{align}
In Fourier space, the Poisson equation has a particularly simple form, and the potential is calculated as \footnote{the $k=0$ mode is set to $0$ and represents a constant density offset. We are uninterested in this mode because a uniform density background does not effect the dynamics.}:
\begin{align}
    \tilde V (k) = -\frac {C}{(2\pi)^2}\frac{\rho'(k)}{k^2} \,.
\end{align}
The potential is then transformed back to configuration space,
\begin{align}
    V(x) = \frac{1}{2\pi} \int  \tilde{V}(k) e^{-ikx}dk \, .
\end{align}
While the above expressions have been written for continuous fields for conceptual clarity, the actual calculations are performed on a grid of size $N_{\rm grid}$ using discrete Fourier transforms.

Finally, the acceleration defined on the grid is computed as 
\begin{equation}
    a(x_i) = a_i = - \sum_j D_{ij}\,V_j\,, 
\end{equation}
where $a(x_i)$ and $V_i \equiv V(x_i)$ are the value of the acceleration and potential at the $x_i$ grid cell respectively. $D_{ij}$ represents the differentiation stencil
\begin{equation}
    D_{ij} = \frac{\delta_{i,j-1} - \delta_{i,j+1}}{2} \,.
\end{equation}
Here $\delta_{i,j} = 1$ when $i=j$ is the Kronecker delta function.
The acceleration at the $i$th particle position, $r_i$, is then interpolated from the grid to the particles position as
\begin{equation}
    a(r_i) = \frac{1}{\Delta x_g} \int a(x) T_i(x) dx \,,
\end{equation}
which has been written as continuous here for conceptual clarity. However in practice, $a(x)$ is defined on the grid. $a(r_i)$ is then the acceleration of the $i$th particle. 
Apart from the computation of the acceleration in the final step, the procedure of going from the density field to the potential as outlined above is also used for all the other solvers that are described below.




\subsection{Schr\"odinger-Poisson solver}
\label{sec:SP}

\subsubsection{Initial conditions}

We generate the initial conditions of this solver by populating our phase space with streams represented by a single one-dimensional (1D) wave function defined on a grid with $N_{\rm grid}$ cells. The wave function can be represented in the following way in either configuration space or momentum space:
\begin{align}
    &\psi(x) = \sqrt{\rho(x)} e^{ i \phi(x) } \, , \\
    &\tilde{\psi}(k) = \sqrt{\tilde{\rho}(k)} e^{i \tilde{\phi}(k)} \,,
\end{align}
where $\psi, \tilde{\psi} \in \mathbb{C}$ are single valued functions of $x$ and $k$, respectively. Recall momentum, $p$, and wave number $k$, are simply related by constants as $p = \hbar k = \tilde{\hbar} u$. To define a stream, we specify a single valued line in either position or momentum space where at each point the amplitude defines the density and the momentum or position is given by the derivative of the phase. For example, a stream $s$ in real space is defined as
\begin{align}
    &\psi_s(x) = \sqrt{\rho(x)} e^{i \phi(x)}\, , \\ 
    &p_s(x) = \hbar \nabla_x \phi(x) \,.
    \label{eq:real_derivative}
\end{align}
The phase should be chosen here such that the quantum pressure of a given stream is small compared to the classical potential. This constraint is discussed in greater detail in Section \ref{sec:Discussion}.

Likewise, a single value stream $\Tilde{s}$ can be defined in momentum space as 
\begin{align}
    &\Tilde{\psi_s}(p) = \sqrt{\Tilde{\rho}(p)} e^{ i \Tilde{\phi}(p)} \,,\\ 
    &x_s(p) = \hbar \nabla_p \Tilde{\phi}(p) \,.
    \label{eq:momentum_derivative}
\end{align}

To build up the equivalent of a velocity dispersion, we add up the contributions of different streams:
\begin{align}
    \psi(x) = \frac{1}{\sqrt{N_s}} \sum^{N_s}_s \psi_s(x)\,.
\end{align}
An individual stream $\psi_s(x)$ with velocity $u_s$, spatial density $\rho_s(x)$, and sinusoidal velocity perturbation with magnitude $\delta u$ and wavelength $\lambda$ can be written as
\begin{equation}
    \psi_s(x) = \rho_s(x) \exp \left( \frac{i u_s x}{\tilde{\hbar}} - \frac{i \lambda \, \delta u}{\tilde{\hbar}} \cos{\frac{2\pi x}{\lambda}} \right) \, .
\end{equation}
We will frequently construct our phase space out a sum of such streams. We define $\tilde{\hbar} \equiv \hbar / m$.

We will discuss below how a sum of streams allows us to approximate warm phase space distributions. It should be noted that formulating the wave function in this way makes phase space periodic in position and momentum by construction. Because the momentum of our wave function is represented by oscillations on the grid the highest momentum that can be simulated before aliasing sets in is 
\begin{align}
    p_{\rm max} = \hbar \pi / \Delta x_g \,,
\end{align}
where $\Delta x_{g}$ is the grid cell spacing. This means the maximum magnitude positive and negative momenta identify since, when defined on a grid, we cannot distinguish left and right going waves with $p_{\rm{max}}$ momentum. 

\subsubsection{Update}
\label{sec:SP_update}

The update rule for the Schr\"odinger-Poisson set of equations follows a scheme similar to the classical update outlined in Sec.~\ref{sec:classical}. It is a spectral kick-drift-kick solver implemented in \cite{MoczBECDM}. We start from the Schr\"odinger equation
\begin{align}
    i \hbar \Dot{\psi}(x) = H \psi(x) =  \left( \frac{p^2}{2m} + mV(x) \right)\psi(x) \,,
\end{align}
where $p \equiv -i\hbar \nabla$. For slowly changing Hamiltonians, the wave function update between time $t$ and $t+\Delta t$ can be approximated as follows \cite{Ball}:
\begin{align}
    \psi_{t+\Delta t}(x) &= \exp \left( \int_{t}^{t+\Delta t} \frac{-iH_{t'}}{\hbar} dt' \right) \psi_{t}(x) \nonumber \\
    &\equiv U(t + \Delta t,t)\psi_{t}(x) \,,
    \label{eq:update_operator}
\end{align}
where $U(T,t)$ is the unitary operator which takes the wave function at time $t$ to the wave function at time $T$. Note that, ``slowly changing" in this context means the time step is chosen such that \cite{Widrow}
\begin{align}
    \frac{(H(t+\Delta t) - H(t)) \, \Delta t}{\hbar} \ll 1 \, .
\end{align}{}
Discretizing this equation we can write the update step that takes $\psi_t \rightarrow \psi_{t+\Delta t}$,

\begin{align}
    \psi_{t+\Delta t} = \exp \left( \frac{-i \, H_t}{\hbar} \Delta t \right) \psi_{t} \,,
\end{align}
which will be accurate when $\frac{H}{\hbar} \Delta t \ll 1$. We also notice that the Hamiltonian is separable into terms which depend only on the momentum or only on position. We can therefore rewrite the Hamiltonian as potential and kinetic terms
\begin{align}
    H &= T(p) + V(x) \, .
\end{align}
This also implies that the operator $U(T,t)$ can also be written as a product of two terms, 

\begin{align}\label{eqn:operators}
    &U_t(\Delta t) \equiv U(t, t+ \Delta t) \nonumber \\
    &= \exp \left( \frac{i \, \Delta t \, p^2}{2m\hbar} \right) \exp \left( \frac{-i \, \Delta t \, m \, V_t}{ \hbar} \right) \exp \left( O( \Delta t^2 [T,V] ) \right) \nonumber \\
    &\approx U^T_t(\Delta t) U^V_t(\Delta t) \,.
\end{align}
Notice that $U^T_t$ is diagonal in the momentum basis and that $U^V_t$ is diagonal in the position basis and correspond to the kinetic and potential energy contributions, respectively. $V_t$ is the potential at time $t$. Also, recall that position and momentum are related by Fourier transform and so the momentum density of our wave function is represented by the k-space wave function scaled appropriately. This means that the update can be separated into two parts the first changes the phase of the configuration space wave function, and the second changes the phase of the momentum space wave function. The update and density deposit rules are described in Table \ref{tab:update} and given below. To compute the potential, we follow the same procedure as for the classical solver. The solver used here is also used in \cite{Garny, Woo_2009}.
\begin{align*}
    \Tilde{\psi}_{t+\Delta t} &= U^T_t(\Delta t/2)\Tilde{\psi}_t \textrm{  (position update half step)} \\
    &\textrm{(calculate $V_t$)} \\ 
    \psi_{t+\Delta t} &= U^V_t(\Delta t)\psi_{t} \textrm{  (momentum update full step)} \\
    \Tilde{\psi}_{t+\Delta t} &= U^T_t(\Delta t/2)\Tilde{\psi}_t \textrm{  (position update half step)} \, .
\end{align*}
To compute the potential, we follow the same procedure as for the classical solver, but with the density defined as $\rho(x) = |\psi(x)|^2$.

The quantum analog of classical phase space density for the Schr\"odinger-Poisson system is described in Sec. \ref{sec:Physics} and Table \ref{tab:update} and is given
\begin{equation}
    f(q,p) = (2\pi \hbar)^{-1}\left| \braket{q,p|\psi} \right|^2 \, .
\end{equation}

It should be noted that in this work the Husimi distribution is only evaluated in post-processing and is not directly factored into the evolution of the system. When using the Husimi as a diagnostic tool, the smoothing length scales should be kept in mind. 

\subsubsection{Formal correspondence to and deviation from classical method} \label{sec:SP_correspondence}

Equations ~\ref{eq:real_derivative} and \ref{eq:momentum_derivative} imply that for a single stream, the derivative of the position space phase can give us the momentum of the stream at a given position, while the derivative of the momentum space phase can give us the position of the stream at a fixed momentum. For a single stream in phase space, at any phase space grid cell populated by a constituent packet of the stream, the position $q$ and momentum $p$ can be defined as 
\begin{align}
    &r \equiv \hbar \nabla_p \Tilde{\phi}(p) \,,\\
    &v \equiv \tilde \hbar \nabla_x \phi(x)\,.
\end{align}
Therefore, the derivative of our phases at that location in phase space gives us the position and momentum of that packet. Further, the amplitude of the wave function at that point gives the density at that $(q,p)$ in phase space. Now recall that our update is as an update of the momentum space and position space phases, respectively, such that we can rewrite our update rule as follows:
\begin{align*}
    \Tilde{\phi}_{t + \Delta t} =& \Tilde{\phi}_{t} + \frac{\Delta t \, p^2}{4m\hbar} \textrm{ (update position half step)} \\
    &\textrm{(calculate V)} \\
    \phi_{t + \Delta t} =& \phi_{t} + \frac{-\Delta t\, mV}{\hbar} \textrm{ (update momentum full step)} \\
    \Tilde{\phi}_{t + \Delta t} =& \Tilde{\phi}_{t} + \frac{\Delta t \, p^2}{4m\hbar} \textrm{ (update position half step)} \, .
\end{align*}
We can then use Equations ~\ref{eq:real_derivative} and \ref{eq:momentum_derivative} to recast the update rules as 
\begin{align*}
    r_{t + \Delta t} =& r_t + \frac{v~\Delta t}{2} \textrm{ (update position half step)} \\
    &\textrm{(calculate V)} \\
    v_{t + \Delta t} =& v_t + a~ \Delta t \textrm{  (update momentum full step)} \\
    r_{t + \Delta t} =& r_t + \frac{v~\Delta t}{2} \textrm{  (update position half step)} \, ,
\end{align*}
where $a \equiv -\nabla_x V$. This set of updates are exactly the same as those used for the classical solver, see equation \ref{cl_update}. At face value, this would suggest that the classical solver and the Schr\"odinger-Poisson solver will agree with each other up to terms proportional to $O(\Delta t^2[T,V])$. However, we assumed an isolated single-valued stream in phase space. It is easy to define the derivative of the phase in either position space or momentum space for a one-dimensional wave function provided there is only one stream. However, this is more challenging when we have multiple streams. Consider the simplest possible case of a system with two Gaussian streams with constant momenta $p_1 = \hbar \nabla_x \phi_1(x)$ and $p_2 = \hbar \nabla_x \phi_2(x)$ and identical position space densities $\rho_i/2$ where $\rho_i(x) \sim \exp{(-x^2)}$. This wave function can be written as a superposition as follows:
\begin{align}
    \psi(x) = \sqrt{\rho_i(x)/2}\left (e^{p_1 xi/ \hbar} + e^{p_2 xi/ \hbar}\right) \,.
\end{align}
For uniform constant momenta we set $\phi_{1/2}$ as follows:
\begin{equation}
    \phi_{1/2} = \frac{\hbar}{p_{1/2}} x
\end{equation}
Using the center of mass momentum $p_c = (p_1 + p_2)/2$ and momentum difference $\Delta p = p_1 - p_2$, we can rewrite the above equation as 
\begin{align}
    \psi(x) = 2\sqrt{\rho_i(x)/2} \cos\left(\Delta p x/2 \hbar\right) e^{i p_c x/ \hbar} \,.
\end{align}
Notice that the amplitude of the wave function oscillates as a function of the position, and depends on the momentum difference $\Delta p$. This stream interference present in the SP method has been noted in \cite{Navy, bertrand_1980, Mocz}. A higher separation in momentum leads to a more rapidly varying function in configuration space as noted in \cite{Navy}. In fact the amplitude of the wave function goes to $0$ wherever the argument of the cosine term is equal to $(n+1)\pi/2$.
It is straightforward to show that the density is also spatially varying,
\begin{align}
    \rho(x) = 2\rho_i(x) \cos^2\bigg(\Delta p \, x/2 \hbar \bigg) \,.
\end{align}
 However, we note that the average value of $\cos^2$ over a period is $1/2$ and so the density is correct as long as we average over a distance $x_{\rm av} = 2\pi \hbar/\Delta p$. This implies that the higher the momentum separation of the streams, $\Delta p$, the smaller the length scale over which the density needs to be smoothed to reproduce the classical solution. Note that this means that the ``classical limit," where the oscillation frequency becomes large, the oscillation present in the wave function remains order unity \cite{Mocz}, however, the fractional distance over which the average of the density approaches the correct value becomes small.
 Because our momenta are represented by a Fourier transform of the spatial wave function, there is an upper limit on the maximum possible momentum representable in the system. This momentum corresponds to the Fourier mode whose wavelength is the same as the grid cell spacing. Thus, our phase space is periodic in momentum space, by which we mean that the maximum and minimum momenta map to each other. Therefore, there is an upper limit on the separation of the two streams along the momentum direction. This suggests that when the averaging length scale set by the momentum difference is smaller than the smallest dynamic length scale of the problem, $x_D$, the classical solution can be recovered. That is, our solution approaches the classical solution when 
\begin{align}
    \frac{x_{av}}{x_D} = \frac{2\pi \hbar}{x_D \Delta p} \ll 1 \,,
\end{align}

\begin{figure*}
	\includegraphics[width = .95\textwidth]{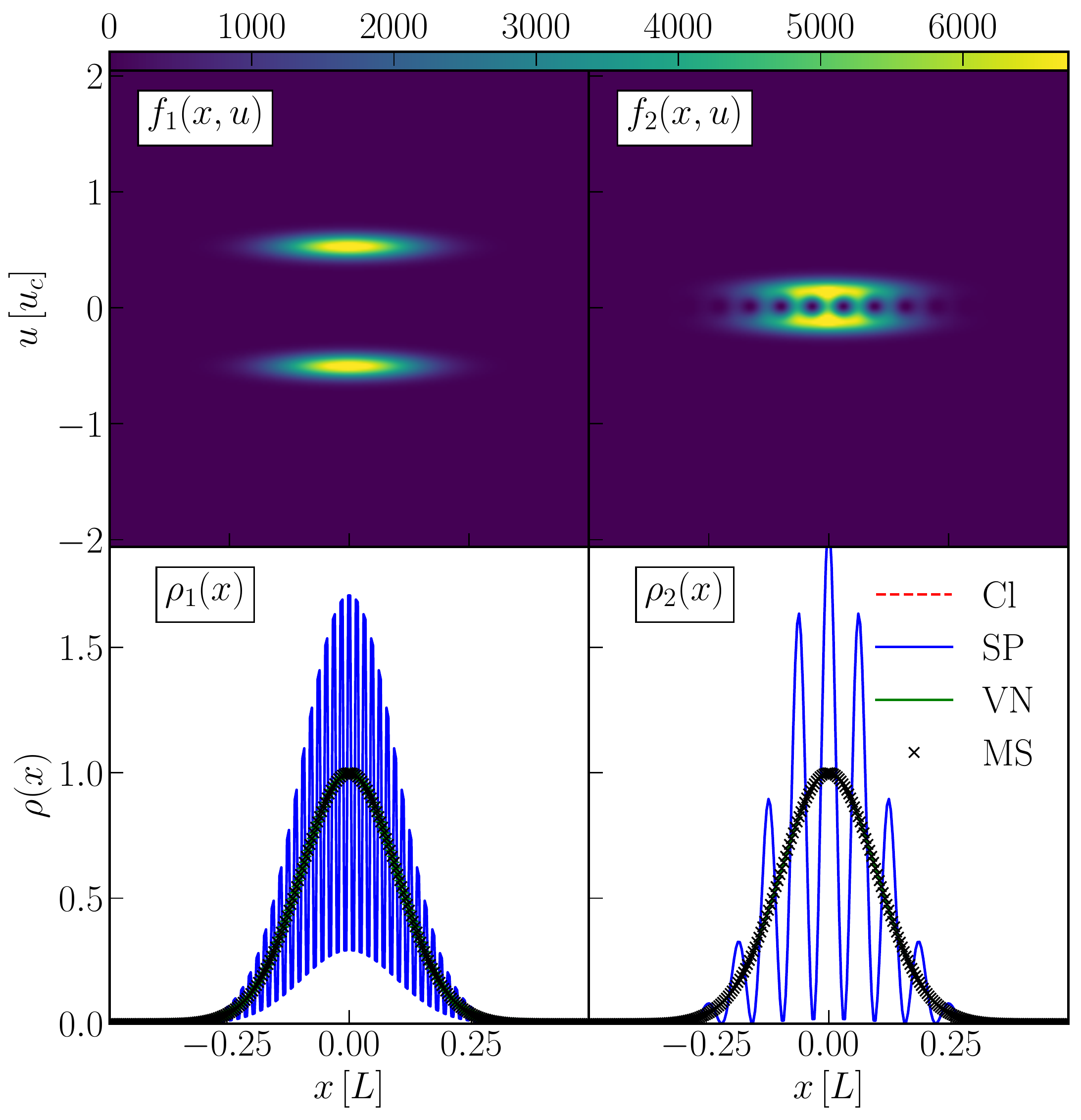}
	\caption{Plot of the Schr{\"o}dinger quantum phase space representation and spatial density of two Gaussian streams. The left column shows the results for well separated streams, and the right column shows poorly separated streams. The wavelength of the oscillation in the Schr{\"o}dinger spatial density decreases as the stream velocity separation increases. The oscillation structure can also be seen in the interference fringes between the stream representations in phase space (top right). The classical (Cl), von Neumann (VN), and multistream/multiple Hilbert spaces (MS) methods lack the oscillation and match the expected density profile.}
	\label{fig:TwoGauss}
\end{figure*}

Fig \ref{fig:TwoGauss} shows our solution deviates strongly from the classical solution when the momentum separation between streams is too small, causing the oscillation term to be of order or greater than the smallest dynamic length scale. 
A similar argument as the one outlined above implies that for a one-dimensional wave function representing two different streams in configuration space, the momentum space resolution has a bound,
\begin{align}
    \frac{p_{\rm av}}{p_D} = \frac{2\pi \hbar}{p_D \Delta x} \ll 1 \,.
\end{align}
It should be noted that it is also possible for a single stream to interfere with itself if appropriately stretched and bent in phase space such that it becomes double valued as a function of $p$ or $x$. 

In the language of quantum mechanics, the implicit assumption contained in the Schr\"odinger-Poisson solver is that the evolved wave function corresponds to a pure state. This means that streams are added in superposition as opposed to being in a classical statistical mixture. Using this method as an alternative to a classical solver, we choose to interpret the wave function densities as corresponding to classical distribution functions but the solver itself allows for wave mechanic effects like interference that would not be present in the classical solution. When using this solver to approximate classical solutions, it should be noted that $\tilde{\hbar}$ is not the physical constant but a simulation parameter qualitatively representing a minimum phase space resolution \cite{Garny}.

It should also be noted that if the spatial smoothing scale of the Husimi distribution, $s \ge \mathcal{O}(x_{\rm av})$, then the interference will not appear in the Husimi representation of the phase space. This property can allow for better classical phase space approximation. However, the phase space agreement can be misleading if the dynamical length scales are also below this length scale, i.e. $s \ge O(x_{D})$,  as the Husimi distribution will not be useful in diagnosing why the field solver and classical solver evolution differ. In this work, the Husimi distribution is only evaluated in postprocessing and is not directly factored into the evolution of the system. This is analogous to constructing a smooth classical phase space density from the N particles.

\subsection{von Neumann-Poisson solver}

As noted in Sec.~\ref{sec:SP}, the Schr\"odinger-Poisson method of simulating classical systems is only valid for initial conditions which can be represented as a pure state. In this section, we present a von Neumann solver that relaxes the assumption contained in the Schr\"odinger-Poisson solver. Specifically, the solver follows the evolution of the density matrix of the system as opposed to a wave function; this allows for the correct evolution of mixed states, as well as pure states. Mixed states correspond to a classical statistical mixture or distribution of wave functions. A classical mixture of wave functions each exists in a separate Hilbert space and thus does not interfere with one another, opposed to quantum superpositions of different streams, where interference terms arise naturally.

The von Neumann scheme increases the numerical scaling of the problem to $O(N^{2n} \log N)$. However, there has been some interest in the solvers with similar scaling \cite{Yoshikawa, ducks, Kormann}, and in some cases where a continuous (as opposed to discretely sampled with sheets) velocity dispersion is necessary to recover the correct solution this solver may be preferable.

\subsubsection{Initial conditions}

The von Neumann method requires the introduction of the density matrix $P$. This is represented as an $N\times N$ matrix in some basis. The diagonal of this matrix is the density in the chosen representation basis. As mentioned earlier, density matrices can represent a classical mixture of quantum states, unlike the wave function \cite{Ball}.

The increased dimensionality of our fundamental object increases the scaling of the solver to be $\mathcal{O}(N^2 \log N)$ for a one-dimensional problem. In this sense, the von Neumann solver is more analogous to solvers which act directly on $2n$ dimensional phase space, where $n$ is the number of spatial dimensions. However, this method may be beneficial as the evolution of this solver still involves unitary operators and so we expect tracking of conserved quantities to remain accurate to machine noise. Additionally, the density matrix still represents a continuous field as opposed to discrete particles. 

To populate the phase space, we can once again use the formalism of adding up multiple streams defined in position space in the following manner with coordinate or index notation, respectively:
\begin{align} \label{density_matrix}
    P(x,x') &\equiv \frac{1}{N_s} \sum^{N_s}_s \psi_s(x) \psi^\dagger_s(x') \,, \\
    P_{ij} &\equiv P(x_i, x_j) \, ,
\end{align}
where $\psi_s \psi_s^\dagger$ represents the outer product of the wave functions, and an individual stream is represented as before,
\begin{align}
    \ket{\psi_s(x)} &= \sqrt{\rho(x)} \exp \left( i \phi(x) \right) \,,\\
    p(x) &= \hbar \nabla_x \phi(x) \,.
\end{align}
Notice that $P$ does not uniquely define a sum of streams, that is, there are infinite possible different summations which would give the same $P$. However, the evolution of $P$ is unique. We index individual elements of the discretized matrix $P$ with $i,j$ such that $P \in \mathbb{C}^{N \times N}$ and $P_{ij} \in \mathbb{C}$. Since the diagonal of $P$ can be treated as a density in the position basis, $P_{ii} \in \mathbb{R}$, i.e. the diagonal elements are real. The density in any basis is the diagonal of the density matrix represented in that basis, 
\begin{align}
    \rho = \textrm{diag}(P) \,.
    \label{eq:VP_density}
\end{align}

We can compute the phase space as 
\begin{widetext}
\begin{align}
    f(p,q) &= (2\pi \hbar)^{-1} \braket{q,p|P|q,p} \\
    &= (2\pi \hbar)^{-1} \int \int dx_i dx_j \braket{q,p|x_i}\braket{x_i|P|x_j} \braket{x_j|q,p} \nonumber \\
    &= (2\pi \hbar)^{-1} (2 \pi s^2)^{-1/2} \int \int dx_i dx_j P_{ij} e^{-((x_i-q)^2 + (x_j-q)^2)/4s^2} e^{-ip(x_j-x_i)/\hbar}  \,. \nonumber
\end{align}

\subsubsection{Update}

The evolution of the density matrix proceeds using the von Neumann equation,
\begin{align}
    P_{t+\Delta t} &=  \frac{1}{N_s} \sum U_t(\Delta t)\psi_{t}\bigg(U_t(\Delta t)\psi_{t}\bigg)^\dagger \\
    &= U_t(\Delta t)P_{t}U_t^\dagger(\Delta t) \, .
\end{align}
Note that update operator $U_t(\Delta t)$ in the above equation is similar in form to the one in Eq.~\ref{eq:update_operator}, and therefore also splits into independent updates of the position space phase and the momentum space phases of the matrix. The update 

\begin{align}
    &\Tilde{P}_{t + \Delta t} = U_t^T(\Delta t/2)U_t^{T\dagger}(\Delta t/2) \Tilde{P}_t \, \textrm{(position update half step)} \nonumber \\
    & \textrm{(calculate $V_t$)} \\
    &P_{t + \Delta t} = U^V_t(\Delta t)U_t^{V\dagger}(\Delta t) P_t \, \textrm{(momentum update full step)} \nonumber \\
    &\Tilde{P}_{t + \Delta t} = U_t^T(\Delta t/2)U_t^{T\dagger}(\Delta t/2) \Tilde{P}_t \, \textrm{(position update half step)} \nonumber
\end{align}

and density deposit rules are described in Table \ref{tab:update}.
\end{widetext}

The potential is computed from the density described in Table \ref{tab:update}. Element wise, the position and momentum update rules can be written as
\begin{align}
    &\Tilde{P}_{ij,t+\Delta t} \,= \Tilde{P}_{ij,t} \exp \left[  \frac{i \, \Delta t}{2 \, m \, \hbar} (p_i^2 - p_j^2) \right] \, , \nonumber \\
    &P_{ij,t+\Delta t} \,= P_{ij,t} \exp \left[  \frac{i \, \Delta t}{\hbar} (V_j - V_i) \right] \, , \\
\end{align}
where $\Tilde{P}$ is the momentum space representation of the density matrix. This is computed by performing a Fourier transform on the axis labeled by $i$ and an inverse Fourier transform on the axis labeled by $j$ of the real space density matrix,
\begin{align}
    \Tilde{P}_{ij} &= \bra{p_i} P \ket{p_j} \nonumber \\
    &= \int \int dx_k dx_l \braket{p_i|x_k}\bra{x_k} P \ket{x_l} \braket{x_l|p_j} \nonumber \\
    &= \int \int dx_k dx_l e^{\frac{i}{\hbar}(p_j x_l - p_i x_k)} P(x_k,x_l) \nonumber  \\
    &=F_{ik}^{-1}F_{jl}P_{kl} \,,
\end{align}
where $F^{-1}_{ik}$, $F_{jl}$ represent the inverse Fourier transform along axis $0$ (horizontal) and Fourier transform along axis $1$ (vertical), respectively.

\subsubsection{Formal correspondence to and deviation from classical method}

For a single stream, the correspondence of the von Neumann method with the classical solution can be shown using the same method as presented in Sec.~\ref{sec:SP_correspondence} the only difference being that we start with the density matrix instead of the wave function. This result is expected since the single-stream scenario represents the equivalent of a pure state in our formalism, and the Schr\"odinger and the von Neumann approach are formally equivalent for pure states \cite{Ball}.

The two Gaussian-streams scenario we discussed in Sec. ~\ref{sec:SP_correspondence}, on the other hand, can be naturally represented as a mixed state in the density matrix setup, see equation \ref{density_matrix}, so that the von Neumann and Schr\"odinger approaches produce different results. To show this, we once again consider two streams with momenta $p_1 \equiv \hbar \nabla_x \phi_1$ and $p_2 \equiv \hbar \nabla_x \phi_2$ and identical position space densities $\rho_i/2$. We write our density matrix as 
\begin{align}
    P_{ij} = \frac{\rho_i(x)}{2} \Bigg[ \exp \left( \frac{i \, p_1}{\hbar} \left(x_i - x_j\right) \right) \nonumber \\+  \exp \left( \frac{i \, p_2}{\hbar} \left(x_i - x_j\right) \right) \Bigg] \,.
\end{align}
The density field is therefore given by 
\begin{align}
    \rho(x) = \textrm{diag}(P) = \frac{\rho_i(x)}{2} \bigg(e^0 + e^0\bigg) = \rho_i(x) \,.
\end{align}
The density field has the correct value as can be seen in Fig \ref{fig:TwoGauss}. The oscillations that were present in the density field in the Schr\"odinger solver for two streams are thus eliminated in the von Neumann method. This implies that different streams  no longer interfere with each another. However, streams can still interfere with themselves if they are evolved to the point where they become highly distorted in phase space. 

\subsection{Multiple Hilbert spaces}
As we saw in Sec.~\ref{sec:SP_correspondence}, interference between different streams leads to the breakdown of the correspondence between the classical solution and the Schr\"odinger-Poisson solution. Since interference terms only arise for wave functions defined on the same Hilbert space, this problem can be mitigated by defining each stream in its own Hilbert space as in \cite{Haas, Schaller}. The individual Hilbert spaces are then evolved in the joint potential. Since the underlying variables of this method are one-dimensional wave functions, rather than the full density matrix, this method has lower computational cost than the von Neumann method, the von Neumann method scales as $\mathcal O(N^2)$, while the Schr\"odinger method scales as $\mathcal O(N)$. Therefore, evolving wave functions in multiple Hilbert spaces scales as $\mathcal O(n_sN)$, where $n_s$ is the number of streams and $N$ is the grid size. When the stream number is small compared to the grid size, this method represents a large speedup over the von Neumann method. 


\subsubsection{Initial conditions}


The object of interest in this method is a vector of wave functions $\Psi$ where 
\begin{align}
    \Psi_i(x) = \frac {1}{\sqrt{N_s}}\psi_i(x) \,.
    \label{eq:rho_ns}
\end{align}
Therefore the $i$-th element in the vector represents the $i$-th stream, with each stream defined the same way as before. The density can now be calculated,
\begin{align}
    \rho(x) = \sum_{i = 1}^{N_s} |\Psi_i(x)|^2 \, .
    \label{eq:n_hilbert_density}
\end{align}

The phase space is given as

\begin{equation}
    f(p,q) = (2\pi \hbar)^{-1} \sum^{n_s}_{i} \left| \braket{q,p|\Psi_i} \right|^2
\end{equation}

\subsubsection{Update}

Each individual stream, $i$, is updated using the same procedure as outlined in Sec.~\ref{sec:SP_update}. The only difference is that the density that enters the Poisson solver is the one described in Table \ref{tab:update} and given below.
\begin{widetext}
\begin{align}
    \Tilde{\psi}_{t+\Delta t,i} &= U^T_t(\Delta t/2)\Tilde{\psi}_{t,i} \textrm{  (position update half step, $\forall i$)} \nonumber \\
    \psi_{t+\Delta t,i} &= U^V_t(\Delta t)\psi_{t,i} \textrm{  (momentum update full step, $\forall i$)} \nonumber \\
    &\textrm{(calculate $V_t$)} \\ 
    \Tilde{\psi}_{t+\Delta t,i} &= U^T_t(\Delta t/2)\Tilde{\psi}_{t,i} \textrm{  (position update half step, $\forall i$)} \nonumber
\end{align}
\end{widetext}

\subsubsection{Formal correspondence to and deviation from classical method}

Again the same definitions of $p$ and $r$ will reproduce the classical update rule as before. Returning to the two stream test problem, we find that our vector can be written with two components as 

\begin{align}
    &\Psi_0(x) = \sqrt{\frac{\rho_i(x)}{2}}  e^{i p_0 x/ \hbar} \, , \nonumber\\
    &\Psi_1(x) = \sqrt{\frac{\rho_i(x)}{2}}  e^{i p_1 x/ \hbar} \, .
\end{align}
Eq.~\ref{eq:rho_ns} implies that the density for this system is given by
\begin{align} 
    \rho(x) &= |\Psi_0(x)|^2 + |\Psi_1(x)|^2 \\ 
    &= \frac{\rho_i(x) + \rho_i(x)}{2} = \rho_i(x) \,.
\end{align}
Therefore, this method recovers the correct density (Fig. \ref{fig:TwoGauss}), that is, without the interference terms which appeared Sec.~\ref{sec:SP_correspondence}.

\section{Test cases} \label{sec:TestCases}
In this section we present the results of simulations of various test cases using the different methods outlined in the previous section. For every test problem, we use the classical solver as a benchmark for the accuracy of the other solvers.

\subsection{Cold stream distributions}

The systems represented here correspond to a classical electrostatically or gravitationally interacting plasma composed of a large number of particles with common mass and charge. The system can be described by a continuous phase space distribution obeying the Vlasov-Poisson equations in which the potential is sourced by the charge density of the particles. When the relevant physics involves a small number of cold streams, well resolved in phase space, all the field solvers are capable of reproducing the classical result. We demonstrate this correspondence by simulating two stream instability and sine wave collapse.

\subsubsection{Two stream instability} \label{test:twoStream}

Our first test problem is 1D two stream instability, a classic instability relevant in the evolution of cold plasma \cite{Infeld_1969, thorne}. For example, this system can describe the evolution of fast electron beams moving through slower moving background solar wind \cite{thorne}. Streams of different velocities with separation below the critical velocity $u_c$ should be unstable and perturbations will grow exponentially.  

This problem demonstrates the ability of all of our solvers to accurately model multiple cold streams well resolved in phase space and recover the expected instability. Two streams at different velocities are given an initial perturbation. We set up the system with the two cold streams spaced apart in phase space such that their velocity separation is less than twice the critical velocity $u_0$ (see Appendix for the definition of $u_0$). The initial perturbation will then grow exponentially for a time following these initial conditions \cite{Anderson, Haas, Infeld_1969}. 

We initialize streams with the dimensionless velocity $u_s \approx \pm 0.8$ (see Appendix for the definition of $u_s$), and introduce a sinusoidal perturbation to the stream velocities with perturbation wavelength equal to the size of the box and amplitude $\delta u = 0.05 u_0$. The amplitude of the perturbation over time can be measured (when the perturbation is approximately linear)following the evolution of the root mean square value of electric field. For these simulations, we use $N = 256$ grid cells, $n = 2^{16}$ particles in the classical solver, $\tilde{\hbar} = 1 \times 10^{-6}$; we define our timescale, $t_0$ and Poisson constant, $C$ such that $C \, t_0^2/4\pi \equiv 1$. 
We expect linear growth in the early evolution, where the electric field is proportional to the amplitude of the spatial perturbation,
\begin{equation}
    E = -\nabla V \sim \delta x \, ,
\end{equation}
where $E$ is the electric field, $V$ is the electric potential, and $\delta x$ is the amplitude of the spatial perturbation.

\begin{figure}
	\includegraphics[width = .45\textwidth]{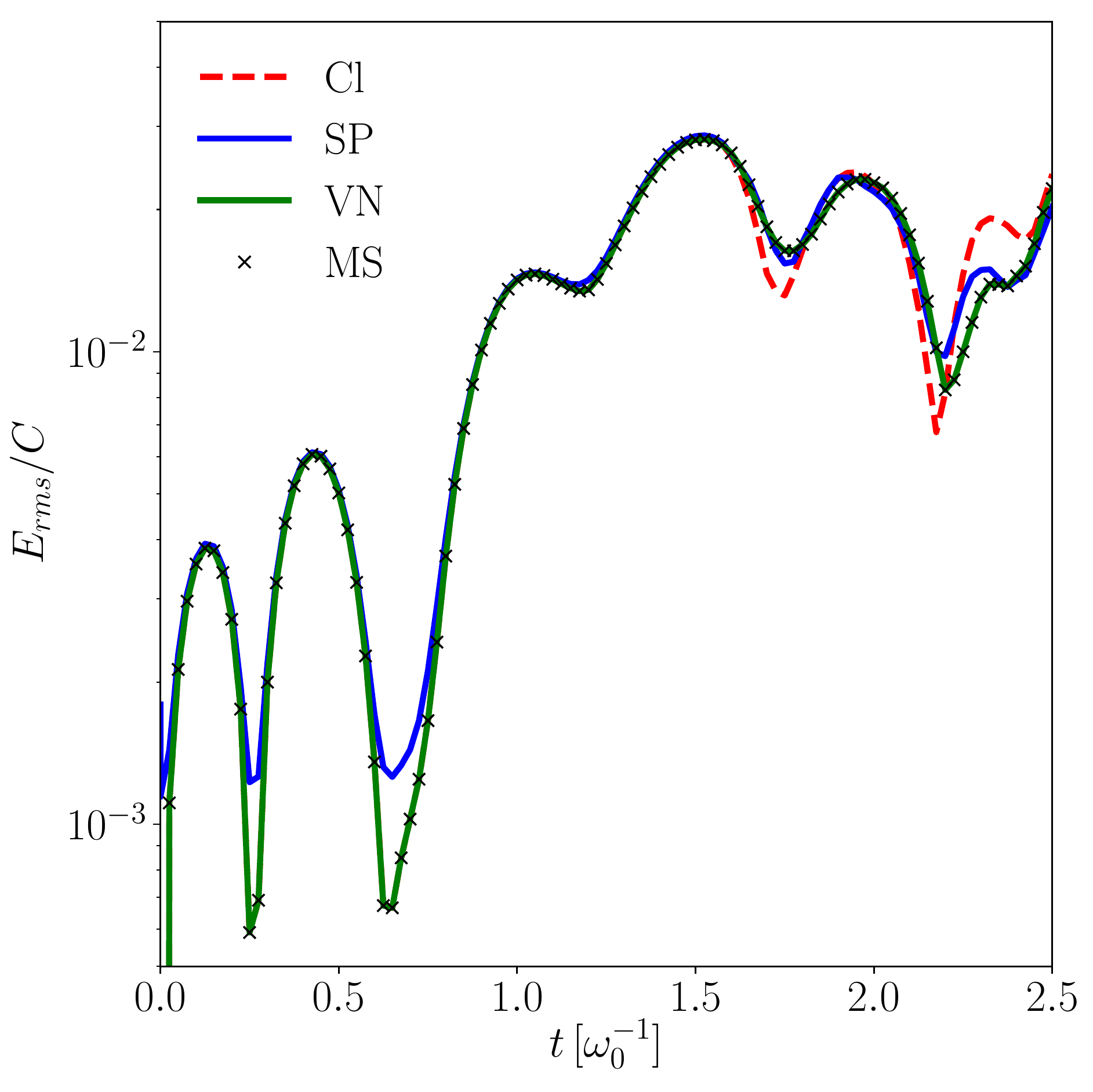}
	\caption{Plotted is the root mean square value of the $E$ field as a function of time for two stream instability simulations for each solver. All four solvers agree quite well even at relatively late times. At early times, when the perturbation is small, we find that interference terms wash our the smallest scale structure in the Schr{\"o}dinger-Poisson (SP) case. However, the solution agrees at later times. Increasing the resolution of these simulations would decrease the smallest resolvable scale for SP solver.}
	\label{fig:TwoStreamE}
\end{figure}

\begin{figure}
	\includegraphics[width = .45\textwidth]{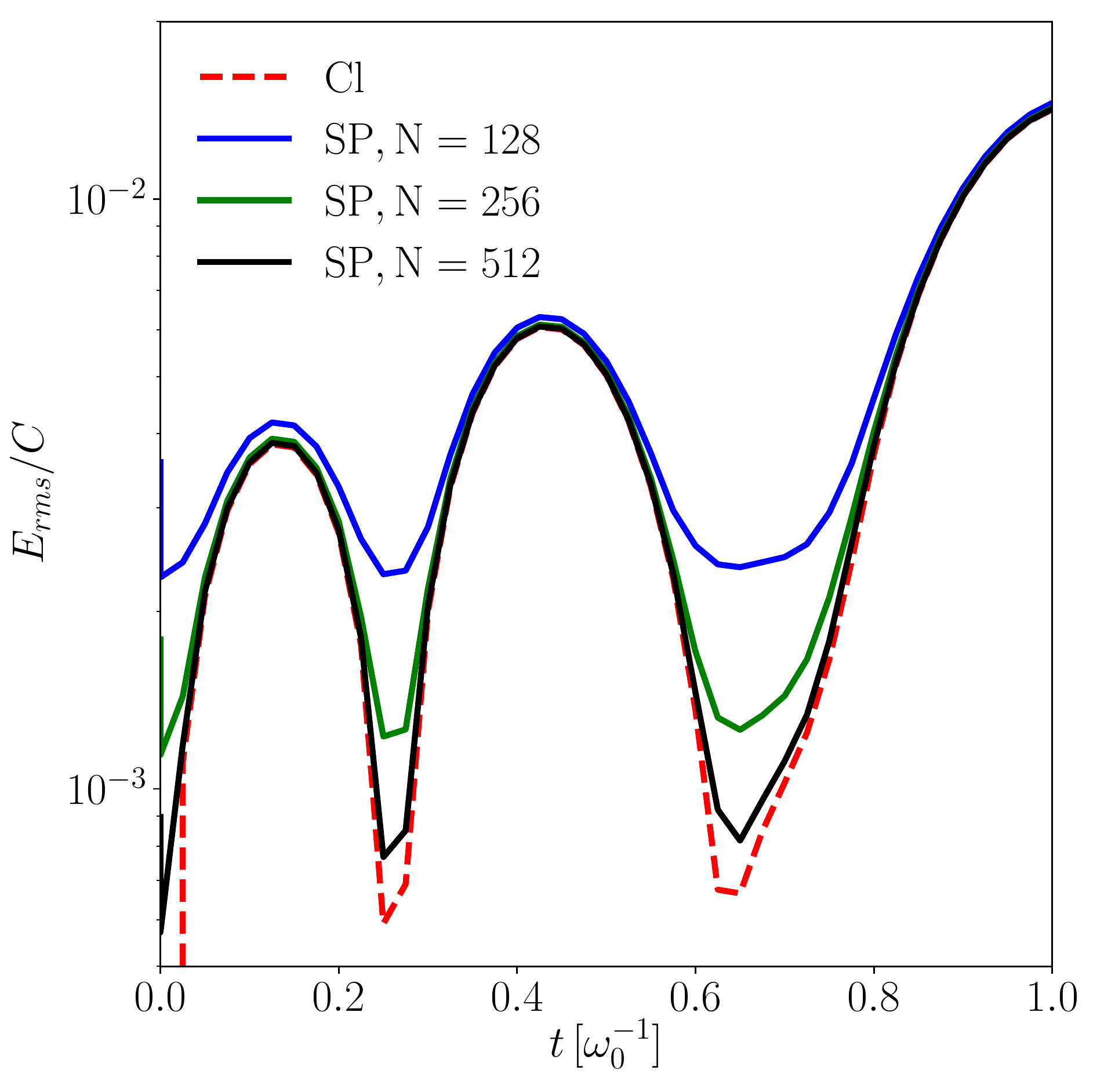}
	\caption{Plotted is the root mean square value of the electric field as a function of time for two stream instability. We plot the results of a classical solver and three Schr{\"o}dinger-Poisson solvers at varying resolutions. The higher resolution simulations have less impactful interference and are able to closer approximate the classical solution especially at small scales where the interference terms will tend to wash out structure.}
	\label{fig:resStudy}
\end{figure}

\begin{figure*}
	\includegraphics[width = .95\textwidth]{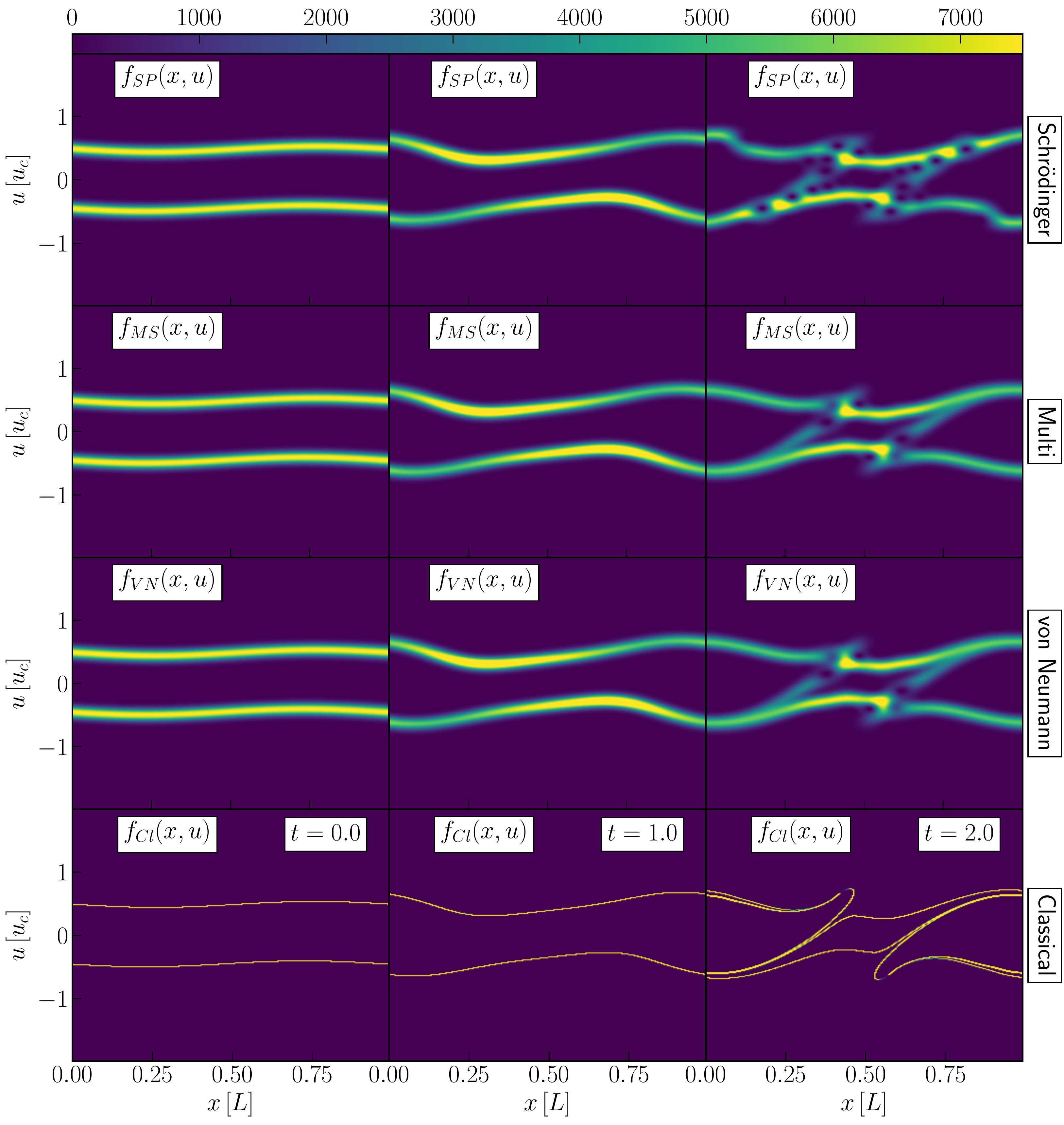}
	\caption{The phase space for two stream instability. Each row represents a different solver and each column a time. All four solvers show roughly the same evolution. The right image in the top row shows interference present in the phase space due to streams overlapping in both position and velocity space at late times. The right images in the middle two rows show interference due to streams interfering with themselves in velocity space for the von Neumann (VN) and multistream (MS) methods. This demonstrates that all methods are effective at representing the evolution of a small number of cold streams for electrostatically interacting systems.}
	\label{fig:TwoStreamPhase}
\end{figure*}

It has been previously demonstrated that classical two stream instability can be reproduced using quantum methods \cite{Anderson, Haas, bertrand_1980}. We find that all four methods produce the same perturbation growth, i.e. all the field solvers agree with the classical solution, as well as with each other. This is shown in Fig.\ref{fig:TwoStreamE}, where we find that the electric field from all the implementations broadly agree. There are some quantitative differences between the Schr\"odinger solver and the other solvers at early times and small length scales, but these disappear with time. The interference terms are responsible for washing out the small scale structure. We can see in Fig. \ref{fig:resStudy} that as we increase the resolution of our simulation, we are able to better approximate the classical solution. We also plot the phase space from each simulation in Fig.~\ref{fig:TwoStreamPhase}, which corroborates the idea that all the solvers produce a similar time evolution, in good agreement with our expectations. Note that all the field solvers, including the Schr\"odinger-Poisson method, are expected to perform well for this test because the separation of the streams is well resolved in phase space. This means that any interference terms in the SP solver to have a vanishing effect on the large scale physics. 

We note that \cite{Haas, Anderson} presented a detailed analysis of the field solver and classical solver correspondence for two stream instability by scanning over two dimensionless parameters. The parts of parameter space which deviated strongly from the classical results are exactly those regions which resulted in the streams being more poorly resolved in phase space compared to the velocity resolution of the simulation. This is in good agreement with the results derived in Sec.~\ref{sec:SP_correspondence}. 

\subsubsection{Sine wave collapse}

We now consider the gravitational collapse of a sine wave perturbation, a standard simulation test problem for self-gravitating collisionless matter. In this problem, the sign of the Poisson constant is made positive to model an attractive potential. The perturbation will grow exponentially resulting in density shell crossings and a characteristic spiral structure in phase space \cite{Abel}. There has been much discussion on the behavior of field solvers during shell crossing \cite{Uhlemann, Garny, Kopp}. When shell crossing occurs, the density and wave function go to zero. Formally, this means that the phase of the wave function at these points is not well defined. Therefore, it is useful to demonstrate the ability of our solvers to reproduce the classical results in systems where shell crossing is expected. 

\begin{figure*}
	\includegraphics[width = \textwidth]{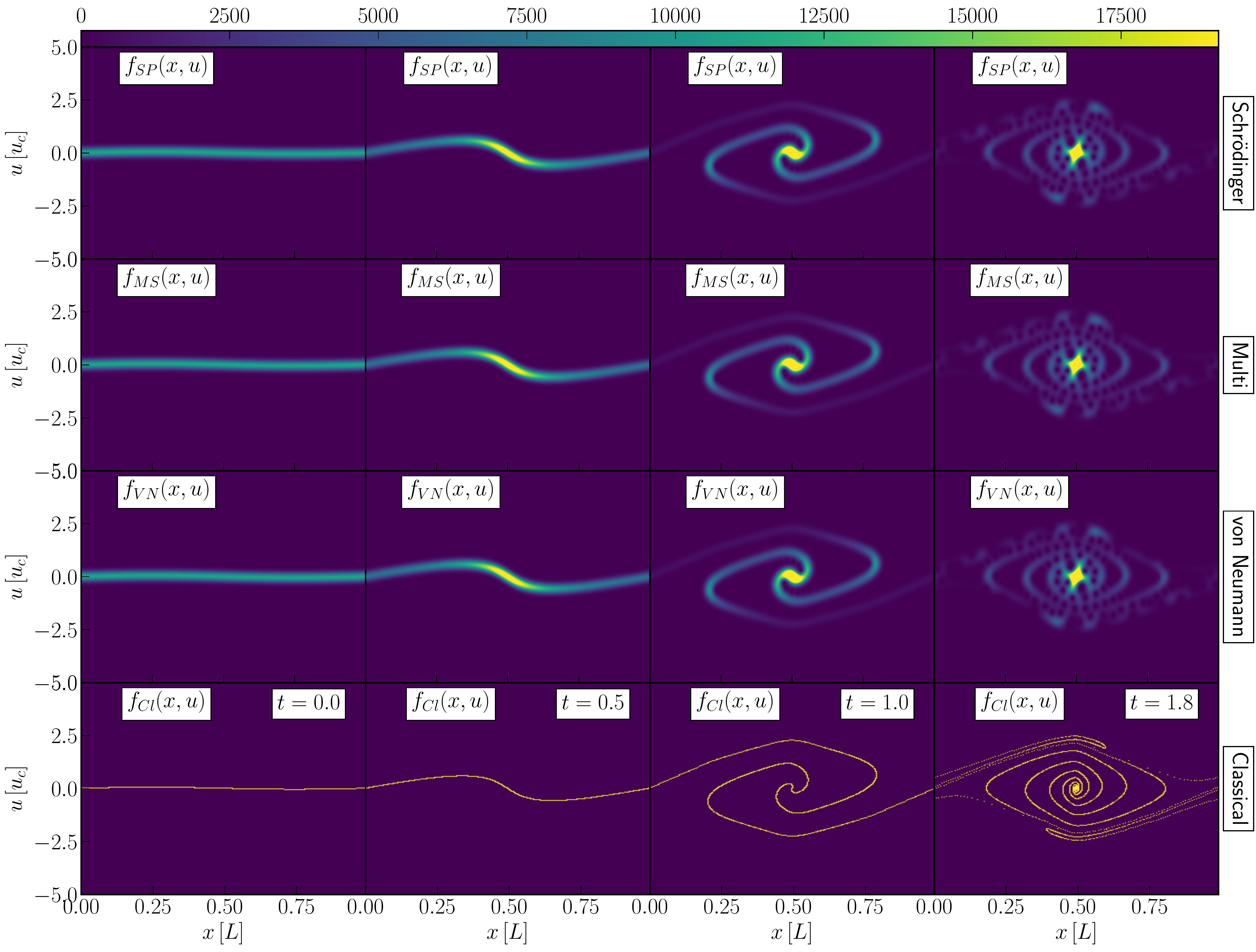}
	\caption{The phase space for 1D sine wave collapse. Each row represents a different solver and each column a time. All four solvers show consistent behavior at early times. This demonstrates that all methods are effective at representing the evolution of cold streams for gravitationally interacting systems. In the final column, we see the field solvers break down as the relevant length scales get smaller as the collapse continues; this is due to the choice of $\tilde{\hbar}$, which sets the simulation phase space resolution.}
	\label{fig:sineWave}
\end{figure*}

\begin{figure*}
	\includegraphics[width = \textwidth]{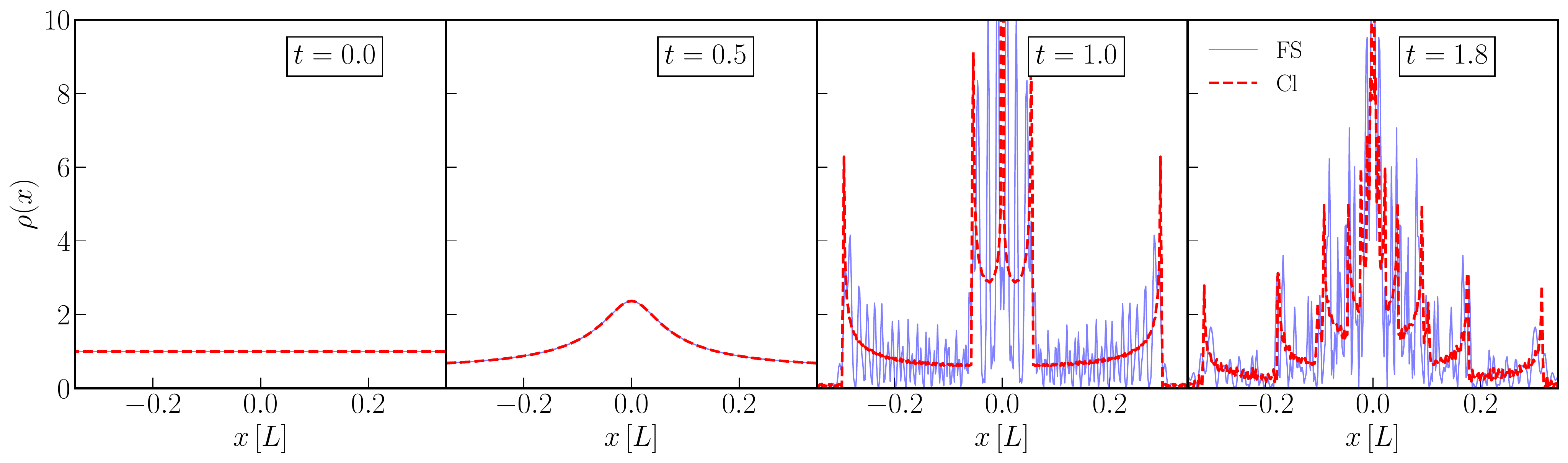}
	\caption{The spatial density for 1D sine wave collapse. All four solvers show consistent behavior initially, but as the stream begins to wind in phase space we see interference terms in the spatial density of the field solvers (FS), here plotted in blue. The evolution of the spatial density for all the field solvers was identical in this test problem, so here we only use the SP solver. At early times, the wavelength of this interference is much smaller than the length scales in the classical system, for example distance between caustics. However, at late times the relevant length scales get increasingly small, eventually approaching the wavelength of the interference.}
	\label{fig:sineWaveDensity}
\end{figure*}

In the one-dimensional case, we initialize a cold stream with a sinusoidal velocity perturbation with wavelength equal to the size of the box and amplitude $\delta u = 0.05 u_0$ and wavelength $L$. For these simulations we use $N = 512$ grid cells, $n = 2^{12}$ particles in the classical solver, $\tilde{\hbar} = 7 \times 10^{-7}$; we define our timescale, $t_0$ and Poisson constant, $C$ such that $C \, t_0^2/4\pi \equiv -1$. We track the phase space of each solver in Fig. \ref{fig:sineWave} and the spatial densities in \ref{fig:sineWaveDensity}.

\begin{figure*}
	\includegraphics[width = .9\textwidth]{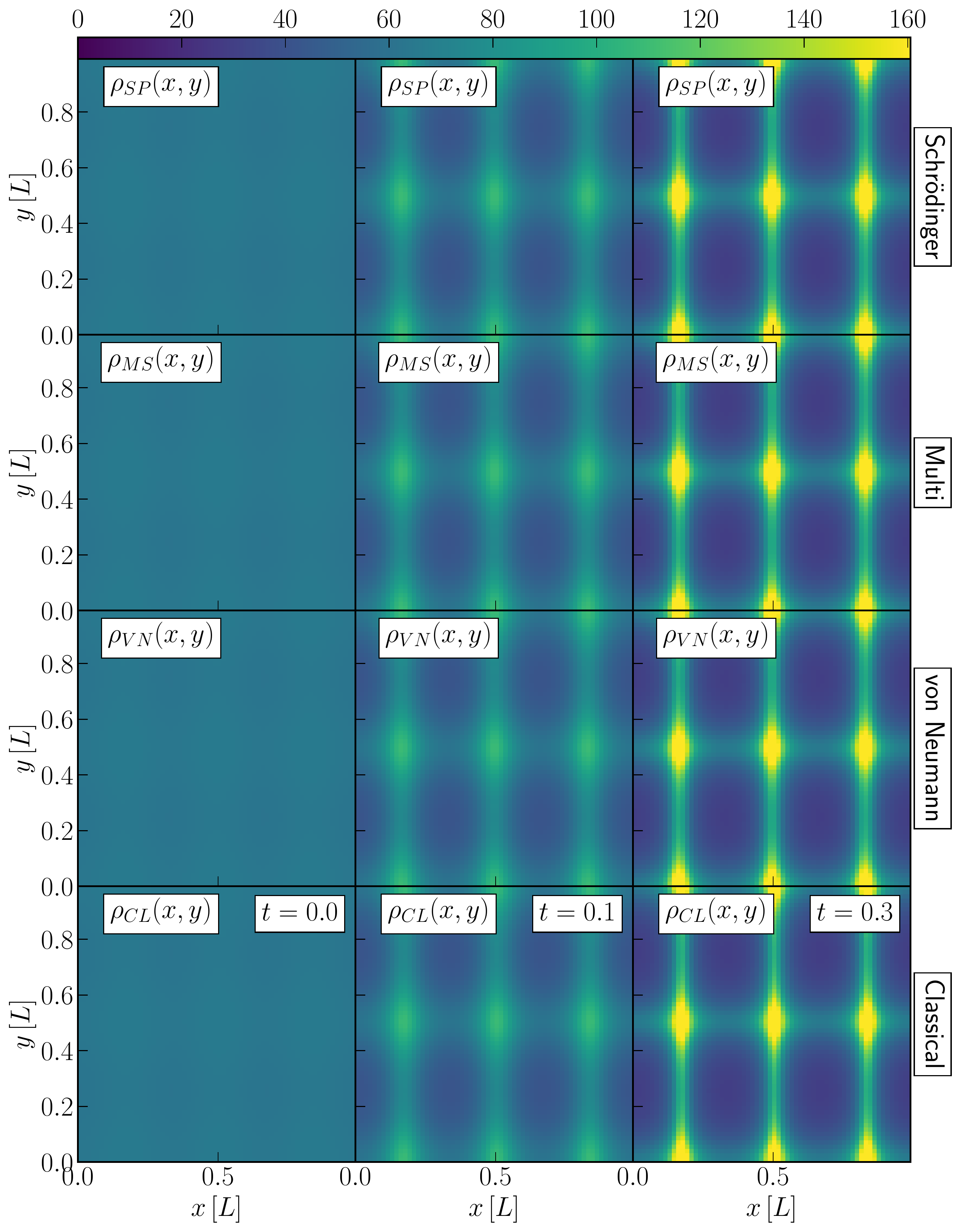}
	\caption{The density for two-dimensional sine wave collapse. Each row represents a different solver and each column a time. All four solvers show the same evolution at early times. This demonstrates that all the field solvers can be extended into higher dimensions and still recover the expected results for the evolution of cold streams.}
	\label{fig:sineWave2D}
\end{figure*}

In the two-dimensional case, we initialize a cold stream with a sinusoidal velocity perturbation in $x$ with wavelength equal to $L/3$ and a velocity perturbation in $y$ equal to $L/2$ both with amplitude $\delta u = 0.1 u_0$. For these simulations, we use $N = 64$ grid cells, $n = 2^{23}$ particles in the classical solver, $\tilde{\hbar} = 7\times 10^{-7}$, we define our timescale, $t_0$ and Poisson constant, $C$ such that $C \, t_0^2/4\pi \equiv -1$. We track the two-dimensional spatial density of each solver in Fig. \ref{fig:sineWave2D}.

As previously demonstrated in \cite{Kopp, Widrow} the field solvers are capable of reproducing the expected results in both one and two dimensions. The phase space in Fig. \ref{fig:sineWave} shows good agreement between all methods. The small scale structure at the center of the spiral at late times is below the Husimi smoothing scale and, as expected, is washed out. However, the large scale spiral structure is reproduced. By sufficiently decreasing the value of $\tilde{\hbar}$ the agreement between the field solvers and classical solvers can be improved in principle, so that the spiral arms and centers would agree well with the classical solution.

\subsection{Warm stream distributions}

The first place that the SP solver begins to fail is for conditions which contain streams that are poorly resolved in phase space. Any conditions with a continuous non-zero velocity dispersion would be considered warm conditions for which stream resolution is not possible. These cases are also of interest to astrophysics and so we demonstrate the ability of the multistream and VN solvers to reproduce the classical results for Landau damping, bump on tail instability, and Jean's instability.   


\subsubsection{Landau damping}

We turn now to the test problem of Landau damping. This effect dampens plasma oscillations by allowing lower energy particles to ``surf" perturbations which transfer energy from the perturbation to the particles \cite{thorne}. We expect initial perturbation amplitudes to exponentially dampen \cite{thorne, Yoshikawa, NGUYEN1981431}. This test problem will demonstrate the ability of our field solvers to represent the interaction of many streams in a warm distribution as well as track the expected decay of perturbations beyond where the classical solver begins to break down.

\begin{figure*}
	\includegraphics[width = .95\textwidth]{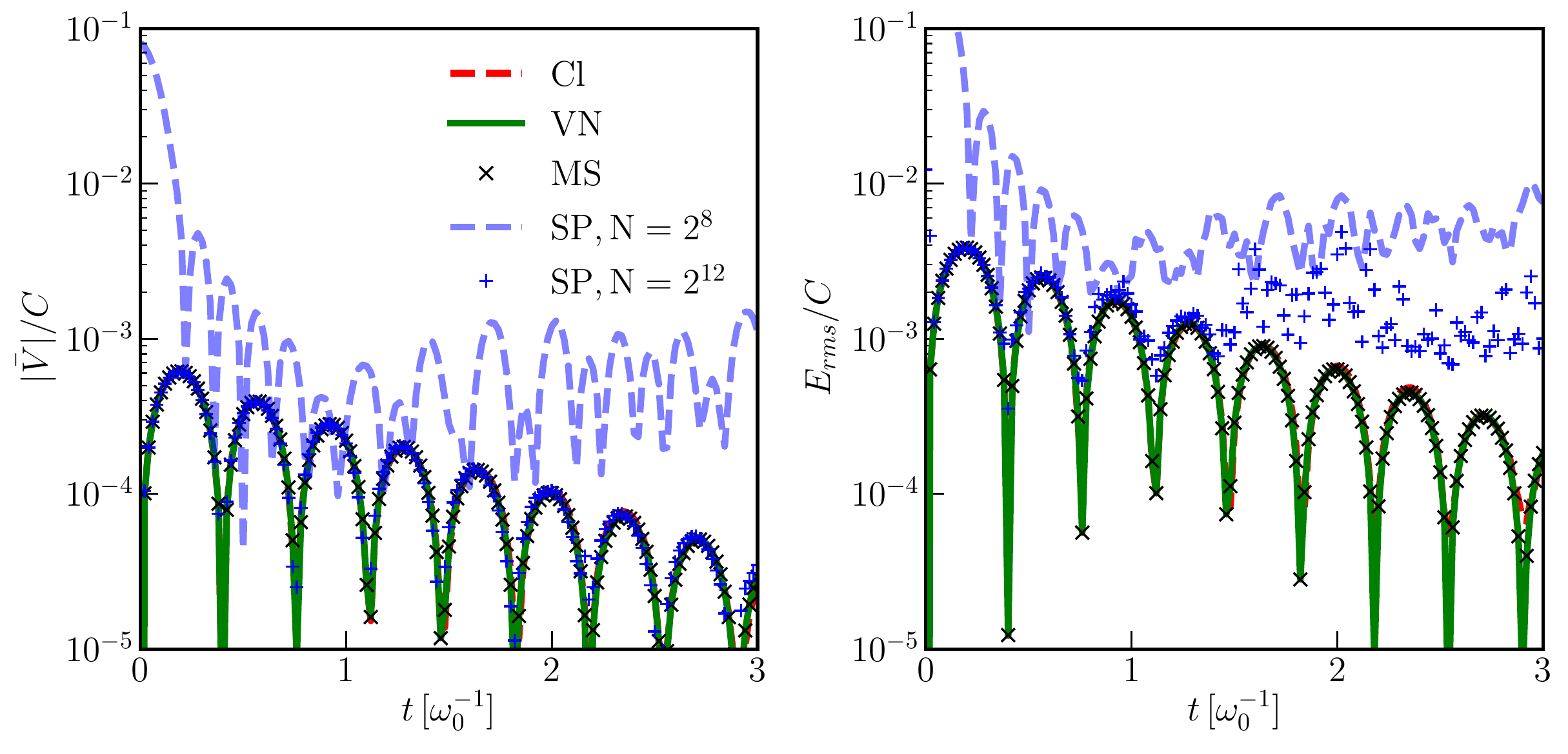}
	\caption{Plotted is the average of the absolute value of the potential (left) and the root mean square value of the electric field (right), as a function of time for Landau damping simulations for each solver. The Schr{\"o}dinger-Poisson (SP) solver fails to reproduce the correct rms $E$ field as the interference terms between the many streams in the system wash out structure below a certain scale. However, if we increase the resolution of our SP solver, to account for the additional computational resources used by the VN and MS solvers, we can recover the correct behavior for the average potential energy although the rms electric field still has a minimum resolvable scale and does not match the other simulations. The higher resolution SP, von Neumann (VN), multistream (MS), and classical (Cl) solvers are in good agreement for the average potential energy, although, the SP is not for the rms electric field. }
	\label{fig:LandauE}
\end{figure*}

Landau damping is driven by the derivative of the dispersion relation \cite{thorne}. Therefore, the correct initial conditions for this problem are a large number of streams approximating a continuous velocity dispersion as opposed to a small number of interacting cold streams which are well resolved in phase space. The initial conditions are a warm stream with Gaussian velocity dispersion with mean $v_{\rm{offset}} = 0$ and standard deviation $v_{\rm th} = 0.49 u_0$. All streams are initialized with velocity perturbation $\delta u = 0.05 u_0$ and wavelength $L$. For these simulations, we use $N = 256$ grid cells, $n = 2^{16}$ particles in the classical solver, $\tilde{\hbar} = 1 \times 10^{-6}$, $N_s = 200$ streams; we define our timescale, $t_0$ and Poisson constant, $C$ such that $C \, t_0^2/4\pi \equiv 1$. We track the amplitude of the density by following the evolution of the root mean square of the electric field over time in Fig. \ref{fig:LandauE}.

The multistream and VN solvers reproduce the solution obtained using the classical solver. The electric field for these methods clearly shows the expected exponential decay. The success of the multistream method was predicted in \cite{Haas}, although it was not demonstrated. The SP solver is unable to reproduce the correct results unless the simulation resolution parameters, $\tilde{\hbar}$ and $N$, are chosen such that streams are well resolved in phase space. When streams are not well resolved in phase space using the SP method, nonphysical interference terms dominate the simulation. It has been shown in \cite{NGUYEN1981431} that an ensembled average of SP solver simulations can produce Landau damping. No such averaging is necessary using the multistream and VN methods. Additionally, a direct integration of the Wigner function on phase space has been shown in \cite{SUH} to reproduce Landau damping. However, as we have demonstrated here, it is not necessary to implement a phase space solver.   

\begin{figure*}
	\includegraphics[width = .95\textwidth]{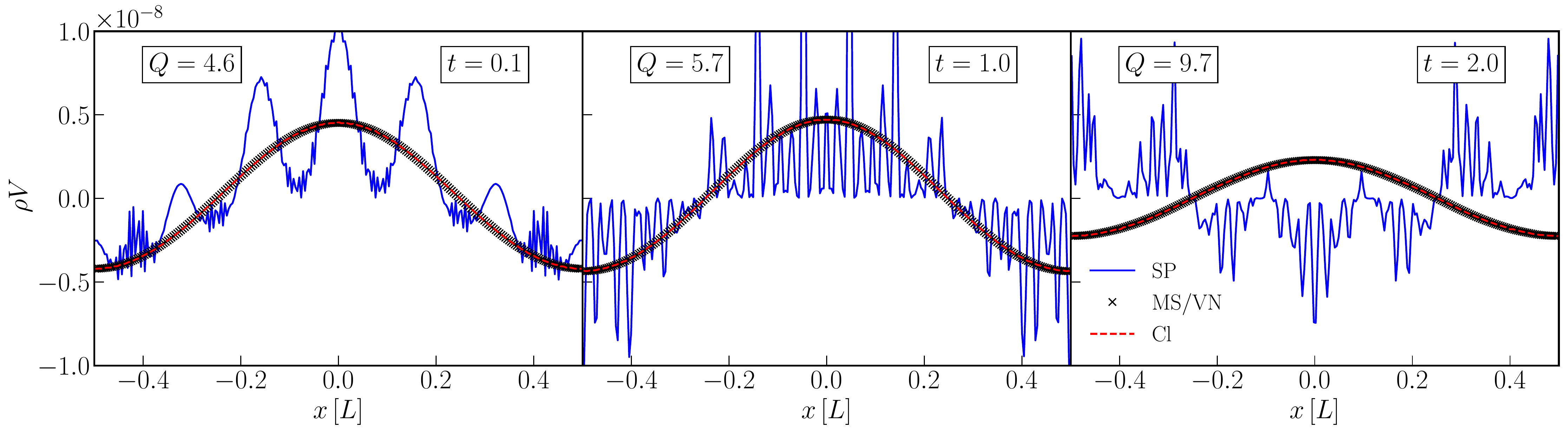}
	\caption{Plotted is the potential for each solver at three different times for a Landau damping simulation using $N_s = 33$ and $N = 2^8$. The MS and VN solvers have identical evolutions in this test problem. Also shown is the average ratio of the density weighted quantum potential to the density weighted classical potential, $Q$. The relatively large number of streams produces interference terms in the SP solver that dominate the evolution and cause its solution to diverge from the classical solution. However, it is unclear from evaluating the quantum potential alone when the solution will diverge.}
	\label{fig:Phimod6}
\end{figure*}

\begin{figure*}
	\includegraphics[width = .95\textwidth]{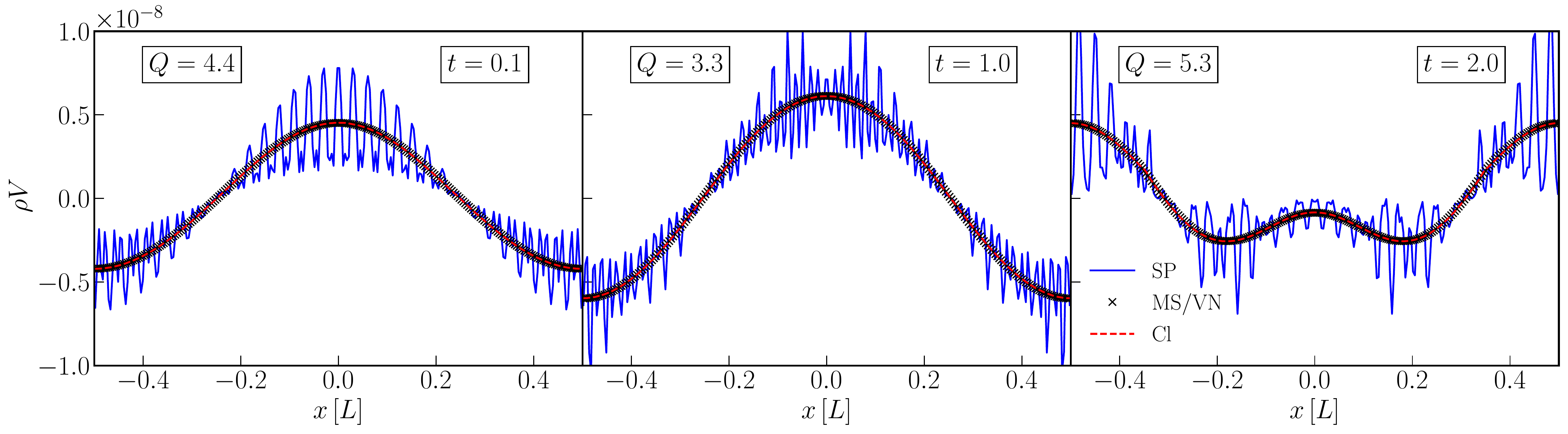}
	\caption{Plotted is the potential for each solver at three different times for a Landau damping simulation using $N_s = 7$. The MS and VN solvers have identical evolutions in this test problem. Also shown is the average ratio of the density weighted quantum potential to the density weighted classical potential, $Q$. With a low number of streams in the system we see that all solvers succeed in recovering the correct behavior, the oscillation is given by the interference discussed in previous sections. Because the oscillation is small compared to the dynamic length scales it has little impact on the large scale behavior of the system. Compared with Figure \ref{fig:Phimod6}, we see that looking at the strength of the quantum potential alone is not enough to indicate strong deviation from the classical solution.}
	\label{fig:Phimod32}
\end{figure*}

The multiple stream interference prevents a simple simulation of Landau damping for the SP solver. If, however, we try and represent the same problem but with a decreasing number of streams we can see how the interference terms affect the evolution by plotting the potential for $N_s = 7$ and $N_s = 33$ in Fig. \ref{fig:Phimod32} and Fig. \ref{fig:Phimod6} respectively. The SP is able to track the solution of the other solvers better when the number of streams is decreased; however, as the number increases the interference terms begin to dominate. 

Also plotted in Fig. \ref{fig:Phimod6} and Fig. \ref{fig:Phimod32} is the spatial average of the ratio of the density weighted quantum potential to the density weighted classical potential term for the SP solver given as \cite{Ball}

\begin{align}
    &V_{QM}(x) = -\frac{\hbar^2}{2m} \frac{\nabla^2 |\psi(x)|}{|\psi(x)|} \\
    &Q \equiv \braket{\rho V_{QM}}/\braket{\rho V} \, .
\end{align}
where $V_{QM}$ is the ratio and $Q$ is the ratio. 

The existence of the quantum potential term when developing the SP system from the Madelung representation is frequently mentioned in the literature \cite{Uhlemann, Garny, Mocz, Haas, Kopp}. However, in the simulations presented in this section, the quantum potential is not a good predictor for when the solution diverges from the classical evolution, as the ratio is of the same order in solutions that both agree, and disagree, with the classical solution. The SP solver still begins to diverge from the correct evolution in Figure \ref{fig:Phimod6} due to the inability to accurately represent the density. It should be noted that a large or small quantum pressure is not necessarily a sign that the SP method will deviate strongly from the classical solution. Other effects such as interference need also be considered. In fact, the quantum potential can reasonably be thought of more as an indicator of velocity dispersion as shown in \cite{Uhlemann}.

\subsubsection{Bump on tail instability}

The bump on tail instability occurs in plasma when a small overdensity at a velocity separated from the background plasma velocity is unstable to perturbations. The importance of the bump on tail instability to plasma dynamics is discussed in \cite{Thurgood, thorne}. Here we use the instability to again demonstrate the ability of our field solvers to produce accurate results for warm stream conditions by correctly tracking velocity density and phase space.

\begin{figure*}
	\includegraphics[width = .95\textwidth]{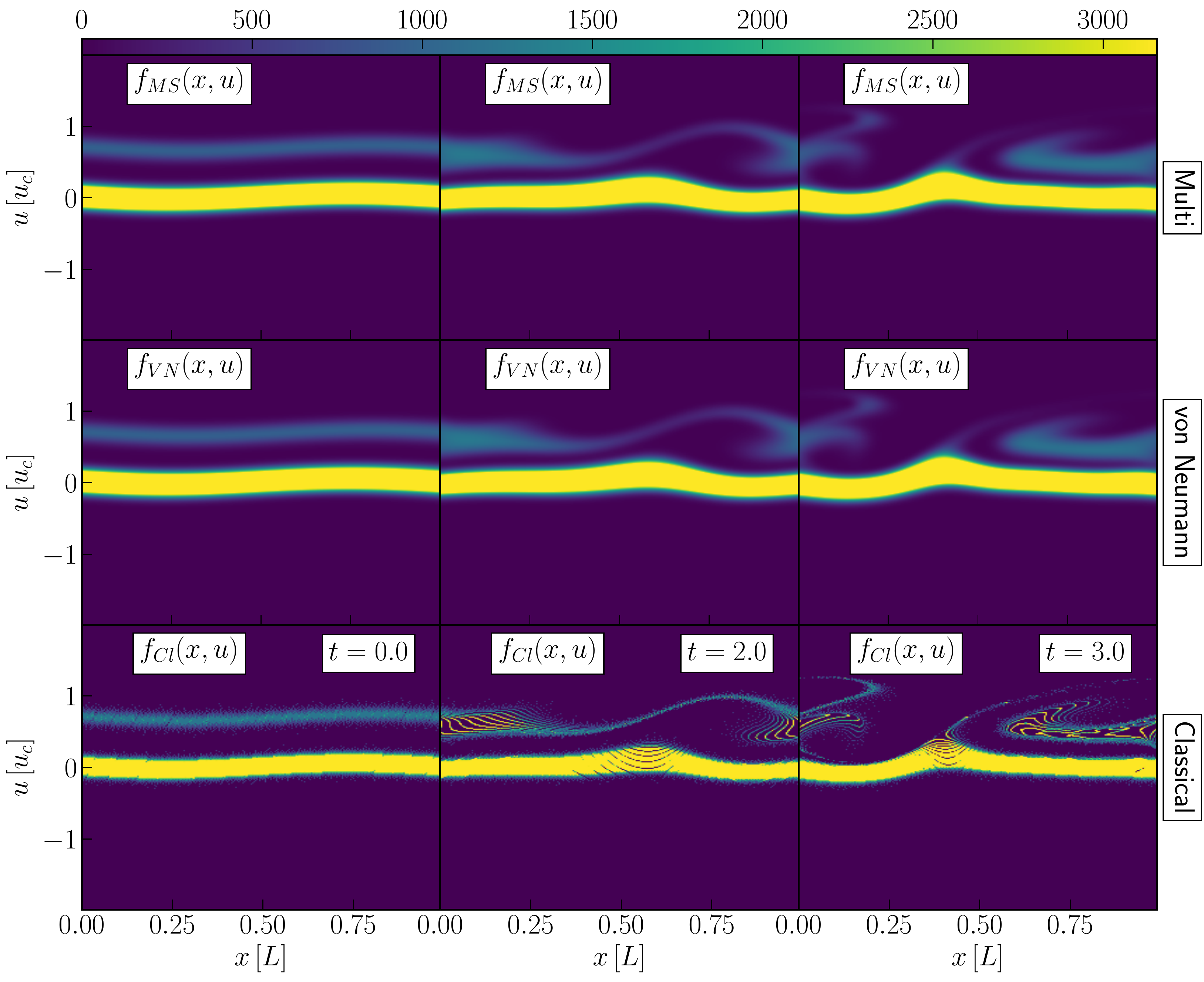}
	\caption{The phase space for bump on tail instability. Each row represents a different solver and each column a time. All three solvers show the same evolution, demonstrating that the field solver extensions can be used to simulate electrostatic systems with a velocity dispersion.}
	\label{fig:BumpPhase}
\end{figure*}

\begin{figure*}
	\includegraphics[width = .95\textwidth]{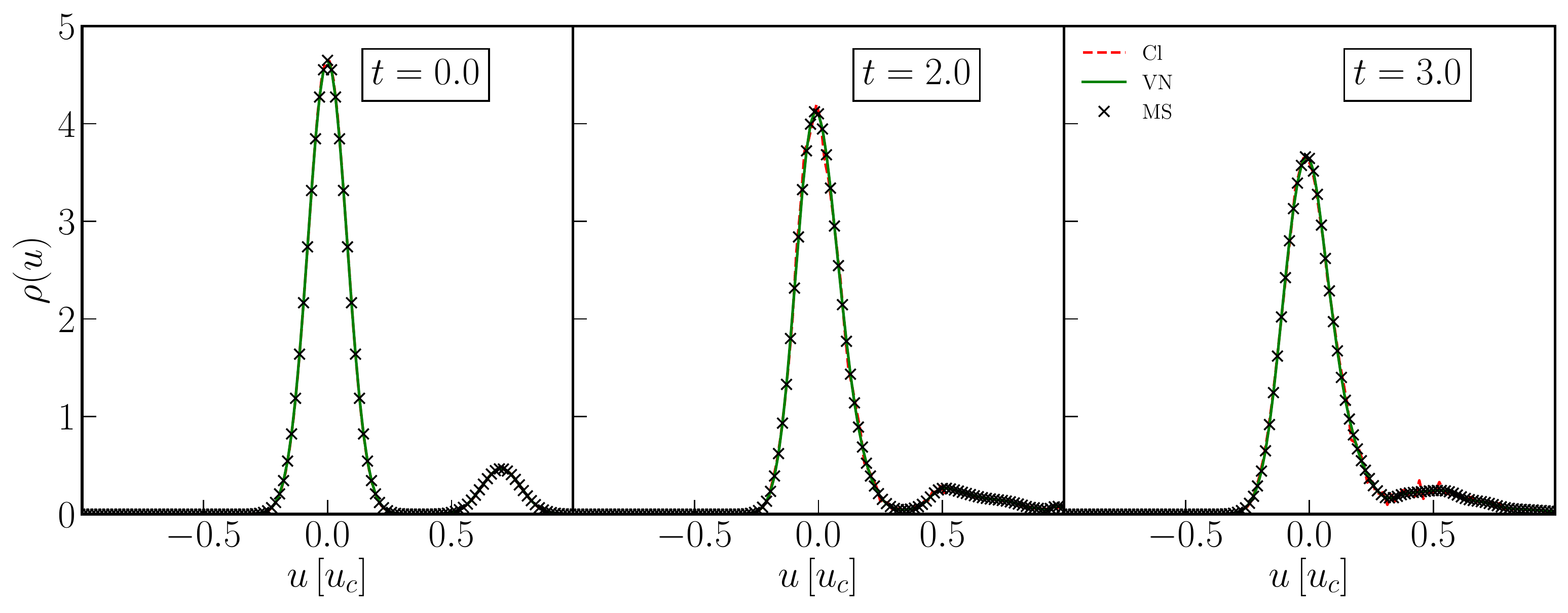}
	\caption{The velocity density over time for each solver for bump on tail instability. All three solvers show the same evolution. Here we see the velocity dispersion represented directly. Even at late times the field solvers are able to very accurately recover the correct distribution without any interference present.}
	\label{fig:BumpRho}
\end{figure*}

Bump on tail instability again involves a continuous velocity dispersion. We start with an initial background distribution with mean velocity $v_{\rm{offset}} = 0$ and thermal velocity $v_{\rm th} = 0.07 u_0$. A bump is created with mass $M_{\rm{bump}}/M_{\rm{background}} = 0.1$ and velocity offset $v_{\rm{bump}} = 0.7 u_0$ and the same velocity dispersion as the background. All streams are initialized with velocity perturbation $\delta u = 0.05 u_0$. For these simulations, we use $N = 256$ grid cells, $n = 2^{16}$ particles in the classical solver, $\tilde{\hbar} = 1 \times 10^{-6}$, $N_s = 200$ streams, we define our timescale, $t_0$ and Poisson constant, $C$ such that $C \, t_0^2/4\pi \equiv 1$. We plot both the phase space in Fig. \ref{fig:BumpPhase} and the velocity density in Fig. \ref{fig:BumpRho} over time. As expected the multistream and VN solvers reproduce the classical results. We see excellent agreement both in the evolution of the phase space and the velocity density. 

\subsubsection{Jean's instability}

For warm distributions subject to their own gravity, the Jean's length determines which modes will be stable or unstable. Perturbations with length greater than the Jean's length will undergo exponential growth while perturbations with length less than the Jean's length will undergo exponential decay \cite{Yoshikawa, thorne}. This test problem will demonstrate the ability of our field solvers to reproduce the correct evolution of the power spectrum at early times for warm initial conditions.

\begin{figure*}
	\includegraphics[width = .95\textwidth]{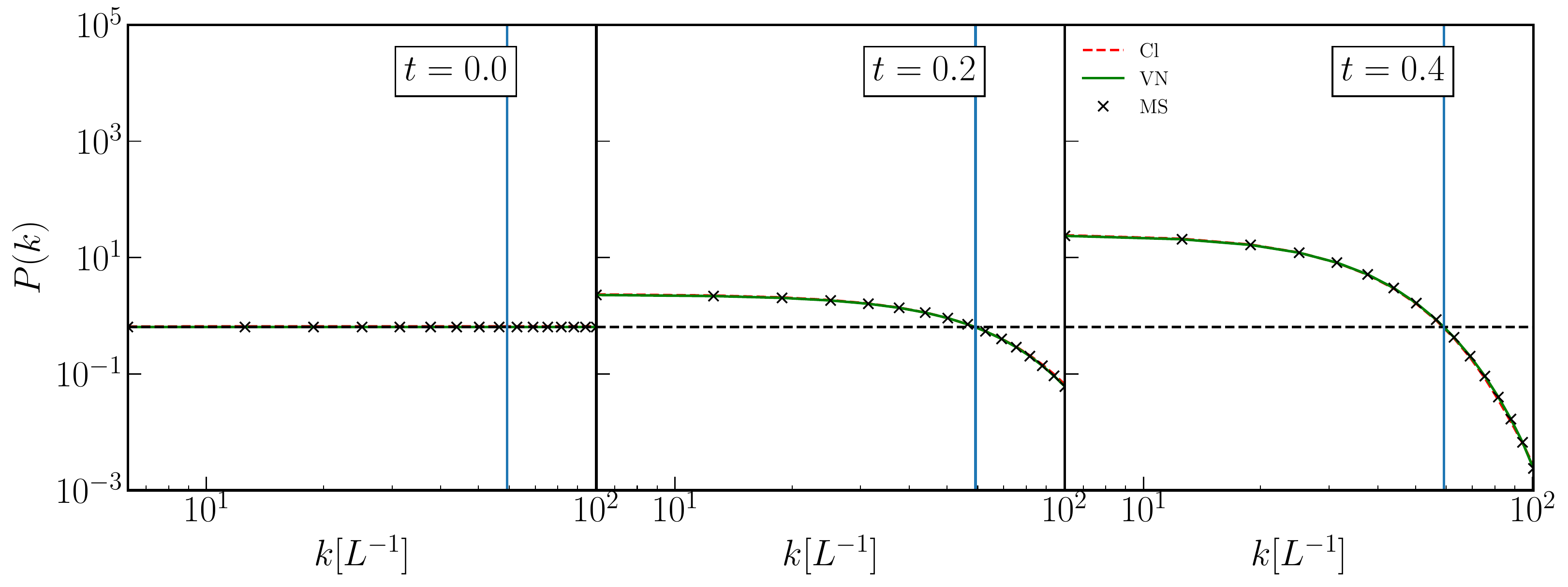}
	\caption{The power spectrum over time for Jean's instability for each solver. The vertical blue line represents the Jean's scale for this simulation, modes to the left should grow and modes to the right should be damped. The horizontal black dotted line shows the initial power spectrum. All three solvers show the same expected growth of modes below the Jean's scale and damping of modes above the scale. This demonstrates that the field solver extensions can be used to simulate gravitational systems with a velocity dispersion as we clearly recover both the large and small scale physics in this system.}
	\label{fig:JeansPower}
\end{figure*}

We start with an initial velocity distribution with $v_{\rm th} = 0.15 u_0$. We then produce a flat initial power spectrum for $k \in [0, 32 \pi / L]$ by adding spatial perturbations with wavelengths $\lambda = [ L/1, L/2 \dots, L/16 ]$ of initial size $\delta_x = .05$. For these simulations, we use $N = 512$ grid cells, $n = 2^{16}$ particles in the classical solver, $\tilde{\hbar} = 1 \times 10^{-7}$, $N_s = 240$ streams, and we define our timescale, $t_0$ and Poisson constant, $C$ such that $C \, t_0^2/4\pi \equiv -1$. We plot the power spectrum over time in Fig. \ref{fig:JeansPower}. 

We can see that the field solvers used here are able to track the classical solver power spectrum. The expected behavior occurs, with modes above the Jean’s scale growing and modes below the scale damping. The von Neumann and multistream solvers successfully reproduce the expected results. Here, like the Landau damping example, we expect that the SP solver is able to reproduce the correct results for small enough $\tilde{\hbar}$.

\section{Discussion} \label{sec:Discussion}

\begin{table*}

\begin{center}

\begin{tabular}{| c | c | c |}

\hline
Cause & Constraint & Parameters \\
 
\hline \hline 

\rule{0pt}{15pt} Kick operator aliasing  & $\Delta t \ll \frac{ \tilde{\hbar}}{V}$  & $\Delta t$  \\[2ex]
\hline

\rule{0pt}{15pt} Drift operator aliasing & $\Delta t \ll \frac{L^2}{\tilde{\hbar}N^2}$ & $\tilde{\hbar}$, $N$, $\Delta t$ \\[2ex]
\hline

\rule{0pt}{15pt} Quantum potential & $\tilde{\hbar} \ll \sqrt{\frac{|\psi_s| \ V_{cl}}{\nabla^2 |\psi_s|}}$ & $\tilde{\hbar}$  \\[2ex]
\hline

\rule{0pt}{15pt} Spatial interference & $\tilde{\hbar} \ll \Delta u \ x_D$ & $\tilde{\hbar}$ \\[2ex]
\hline

\rule{0pt}{15pt} Momentum interference & $\tilde{\hbar} \ll \Delta x \ u_D$ & $\tilde{\hbar}$ \\[2ex]
\hline

\rule{0pt}{15pt} Momentum aliasing & $N > \frac{u_{ph} L}{ \tilde{\hbar} \pi}$ & $\tilde{\hbar}$, $N$ \\[2ex]
\hline

\end{tabular}

\caption{\label{tab:contr} A list of the constraints on simulation parameters necessary for field solvers to approximate classical systems. The first column describes the cause of the constraint, which is listed in the second column. The third column lists which simulation parameters are relevant to the constraint in question. The spatial interference constraint only effects streams in the same Hilbert space and therefore, for warm distributions before nonlinearity, this constraint does not apply to the von Neumann and multistream solvers. }

\end{center}

\end{table*}


\subsection{Interference}

The most obvious problem with field solvers is interference between streams that are poorly resolved in phase space. The interference term comes from the inability of a single valued function of momentum or position to represent a generic function in phase space, for instance, multiple streams that are too closely spaced. This is most problematic for the Schr\"odinger-Poisson solver with only a single wave function. The interference between overlapping streams can be overcome by separating the Hilbert spaces of the streams by either putting them into a classical mixture as in the von Neumann solver or by separately evolving each stream in a shared potential. However, it should be noted that this does not prevent streams from interfering with themselves if a single stream becomes multivalued in either position or momentum. 

Interference can have a large effect on the ability of solvers to approximate the classical solution. If the interference occurs over length or momentum scales that are relevant to the evolution of the system, the solver may deviate strongly from the classical solution. This leads to the two interference constraints listed in Table \ref{tab:contr}. It should also be noted that interference can prevent a solver from being accurate even where the quantum pressure term is vanishingly small. For example, if you are interested in length scales smaller than the oscillation. Therefore it is not necessary to demand a vanishingly small quantum pressure in order to reproduce the correct classical results. 

Recall that the relevant quantity governing the interference is proportional to $\tilde{\hbar}$. This is analogous to the physical constant, but in these simulations is just a free parameter governing the relationship between the numerical representations of position and momentum. Therefore, we can arbitrarily decrease the interference between streams by simply lowering the value of $\tilde{\hbar}$. This of course incurs another problem discussed in the next section.

\subsection{Fixed maxima and aliasing}

The momentum and position representations of the wave function encode information about density in phase space by using derivatives of the complex angle in the wave function. The maximum wave number that can be accommodated by a finite grid of size $N$ and physical length $L$ is

\begin{equation} \label{k_max}
    k_{\rm max} = N \pi / L \,.
\end{equation}

\begin{figure*}
	\includegraphics[width = .95\textwidth]{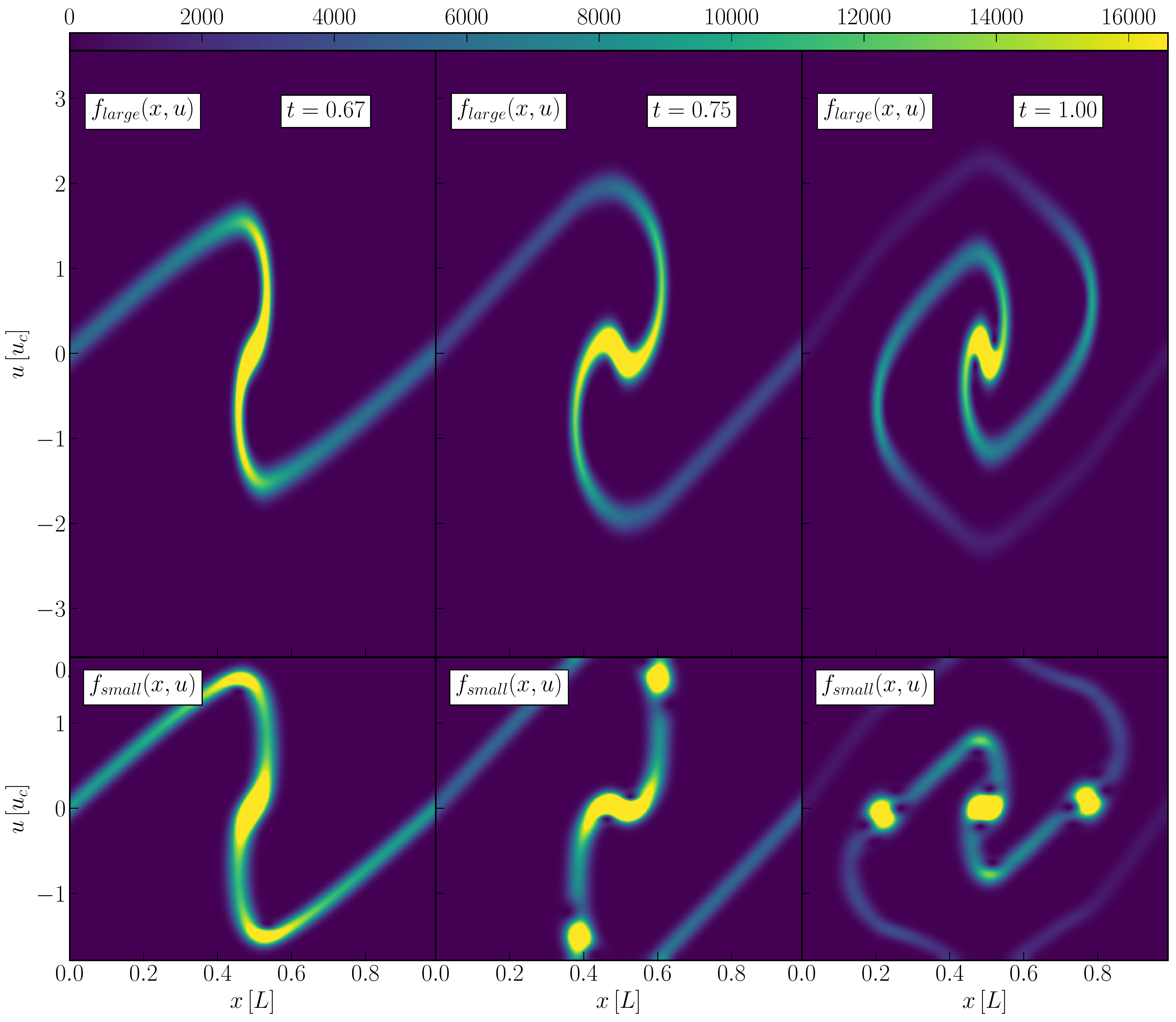}
	\caption{Shown is the phase space for sine wave collapse with $\tilde{\hbar} = 5 \times 10^{-7}$ with $N = 512$ and $N=256$ grid cells for the top and bottom rows respectively. Between $t=0.67$ and $t=0.75$, the simulation first attempts to represent velocities beyond the maximum allowed velocity in the smaller simulation. The bottom row shows how this creates spurious overdensities in phase space that persists in the simulation. Explicit checks need to be performed to ensure that overdensities due to aliasing are not created.}
	\label{fig:alias}
\end{figure*}

Momentum and wave number are related by $p = \hbar k$, implying that the maximum physical momentum that can be represented in the system is fixed by the starting grid size and choice of $\hbar$ and $L$. Consequently, while a wider range of physical momenta can be achieved by increasing $\hbar$, this also increases the influence of interference terms. This means the classical limit is only achieved by making $N$ large and not by varying the value of $\hbar$. The fact that position and momentum are related by the Fourier transform on a finite grid means that the phase space is periodic in both momentum and position space. Therefore, while periodic positions are generally not problematic, the maximum momentum present in the simulation should be checked against the maximum allowed momentum to prevent the formation of spurious structures as in Fig. \ref{fig:alias}. This leads to the momentum aliasing constraint listed in Table \ref{tab:contr}.

\subsection{Taking $\hbar \rightarrow 0$}

Generally, when we want the limit in which a quantum system becomes classical we take the limit $\hbar \rightarrow  0$. However, it is not immediately clear what this limit is taken with respect to. It becomes increasingly confusing when $\hbar$ represents not a physical constant but a simulation parameter, as when using field solvers for classical problems. 

It is clear from the Madelung formalism, discussed in other works \cite{Uhlemann, Garny, Mocz, Haas, Kopp}, that for cold initial conditions $\hbar$ should be sufficiently small such that the quantum potential term is small compared to the classical potential. i.e. $\tilde{\hbar} = \hbar/m$ must be chosen such that for each stream

\begin{equation} \label{V_Q_constr}
    \tilde{\hbar} \ll \sqrt{ \frac{|\psi_s| \ V_{cl}}{\nabla^2 |\psi_s|} } \, .
\end{equation}

However, we have shown that the above condition is insufficient to assure that the classical solution is obtained. And, in fact, we expect a non-negligible quantum potential for warm distributions. This constraint is listed in Table \ref{tab:contr}. We must also consider the interference terms in the representation of the density. Let us again consider two streams separated by $\Delta p = m \Delta u$ in phase space. As we take $\hbar \rightarrow 0$, the density will never approach the classical density. Instead the frequency of the oscillation will increase, meaning a smaller and smaller spatial average will be needed to approach the classical solution. The distance over which it is necessary to average to achieve the classical density must be small compared to the dynamical length and velocity scales of the system $x_D$ and $u_D$. By taking the necessary spatial average to be the order of the wavelength of the interference we can say that 

\begin{align} \label{contr_hbar}
    \tilde{\hbar} \ll \Delta u \ x_D \\ 
    \tilde{\hbar} \ll \Delta x \ u_D \nonumber
\end{align}
must also be satisfied, which is our interference constraints in Table \ref{tab:contr}. The similarity to the uncertainty principle is not a coincidence. $\Delta x$ refers to the distance between over densities in phase space at a constant velocity. Qualitatively, $\tilde{\hbar}$ can be thought of as an area in phase space. Multiple streams sharing a phase space region on the order of $\tilde \hbar$ will begin to deviate from the classical solution. Streams and overdensities can also interfere if they overlap in momentum not just position. Note that this implies that in the nonlinear regime a single stream can interfere with itself. Thus, we also need to consider the dynamical velocities, $u_D$, and spatial separation of phase space over densities at these velocity, $\Delta x$. However, the appropriate orientation of this phase space area depends on the relevant $x_D$ at a given point and so this interpretation of $\tilde \hbar$ should be used conservatively. 

\subsection{Classical approximation}

Using field solvers as approximations of classical systems can be understood as satisfying a series of requirements on the simulation parameters. Parameters must be chosen to address each of the following

\begin{enumerate}
    \item Operator aliasing
    \item Quantum potential term
    \item Wave function interference
    \item Momentum aliasing
\end{enumerate}

Temporal aliasing refers to the fact that we evolve our system with unitary operators represented as arrays of complex numbers. Any given complex number $z = A e^{i\phi}$ does not have a unique representation. We can write

\begin{equation}
    z =  A e^{i\phi} \equiv  A e^{i(\phi + 2 \pi m)}
\end{equation}
for $m \in \mathbb{Z}$. As noted in \cite{Garny, MoczBECDM}, we must choose a time step such that our unitary operators which evolve the system forward in time, $U = U^VU^T =  e^{i \Delta \phi(x)} e^{i \Delta \phi(k)}$ have phases all small compared to unity, i.e.

\begin{equation}
    \Delta \phi \ll 1 \, .
\end{equation}

Using equation \ref{eqn:operators}, implies two constraints on our time step, $\Delta t$,

\begin{align}
    \Delta t &\ll \frac{m \hbar}{p^2_{max}} \\
    \Delta t &\ll \frac{\hbar }{m V_{max}}
\end{align}

We can then substitute in our maximum representable momentum using equation \ref{k_max}. In general, the maximum possible potential should be proportional to the maximum density which means roughly it should scale as $V_{\rm max} \propto M_{\rm tot}N^d/L^{d-2}$ where $d$ is the number of spatial dimensions in the system. Therefore we can rewrite our constraints on the time step and simulation resolution, $N$, as

\begin{align}
    \Delta t &\ll \frac{L^2}{\tilde{\hbar}N^2} \\
    \Delta t &\ll \frac{ \tilde{\hbar} L^{d-2} }{M_{\rm tot}N^d} \, .
\end{align}

$L$ and $M_{\rm tot}$ can generally be set to $1$.

The quantum potential is caused by an inability to compress quantum states to occupy areas in phase space smaller than $\hbar$. The effects of the quantum potential on the use of field solvers as classical approximations has been well studied \cite{Uhlemann, Garny, Kopp, Haas, Kopp} and results in the constraint on $\tilde{\hbar}$ given in equation \ref{V_Q_constr}

\begin{equation}
    \tilde{\hbar} \ll \sqrt{ \frac{|\psi_s| \ V_{cl}}{\nabla^2 |\psi_s|}} \nonumber \, .
\end{equation}

Interference between streams represented in the same Hilbert space can also prevent simulations from being accurate approximations if the length scale of the interference approaches the dynamical length and velocities scales of the system being simulated. Taking this into consideration, we arrive at the constraint given in equation \ref{contr_hbar}.

\begin{align} 
    \tilde{\hbar} \ll \Delta u \ x_D \nonumber \\ 
    \tilde{\hbar} \ll \Delta x \ u_D \nonumber \, .
\end{align}

The first inequality is the spatial interference constraint which only effects streams in the same Hilbert space and therefore, for warm distributions before nonlinearity sets in, this constraint does not apply to the von Neumann and multistream solvers. It should also be noted that it is not just the actual de Broglie wavelength associated with particles at stream velocities $u_s$ we are interested in for interference terms but the de Broglie wavelength associated with the difference in stream velocities $\Delta u$. 

We have shown that spatial interference in certain limits causes the breakdown of the Schr\"odinger-Poisson system as an approximation of a classical system and that this can be particularly problematic when working with warm distributions that include many streams. The wave function interference limits are extremely important to satisfy when trying to reproduce classical evolution and often provide a more relevant scale to compare with $\hbar$ than limitations set by the quantum pressure. The von Neumann and multistream solver extensions we presented here are able to simulate warm stream physics because they put each stream in its own Hilbert space which prevents spatial interference terms.

Momentum aliasing is caused by the fact that our momentum states are represented by waves on a spatial grid. This means that the positive and negative maximum magnitude momentum identify with each other. The maximum physical momentum in the system $p_{\rm ph} = \hbar k_{\rm ph} = m u_{\rm ph}$ must remain below the maximum representable momentum. This gives the following constraint on the simulation resolution, $N$, and $\tilde{\hbar}$.

\begin{equation}
    N > \frac{u_{\rm ph} L}{ \tilde{\hbar} \pi}
\end{equation}

The combination of these constraints prescribes a choice of simulation parameters $\Delta t$, $\tilde{\hbar}$, and $N$, necessary to approximate a classical system with properties $u_{\rm ph}$, $x_D$, $u_D$. These constraints are summarized in Table \ref{tab:contr}.



\section{Conclusions} \label{sec:Conclusion}

We have demonstrated that field solvers can be used to approximate the classical evolution of a number of plasma and gravitational systems. Single wave function field solvers, such as our Schr\"odinger-Poisson solver, can simulate initial conditions in which simulation parameters are chosen such that streams are well resolved in phase space. Furthermore, we have shown that introducing additional wave functions, as in our multiple Hilbert solver, or multistream solver, we can extend the region of validity of field solvers to include warm initial conditions where streams are very close in phase space. We have also introduced a new extension using the density matrix formalism and von Neumann equation. This solver is somewhat analogous to phase space solvers in that it is a $2n$ dimensional solver, where $n$ is the number of dimensions. We have demonstrated that this solver can also be used to simulate warm distributions. 

The limits in which field solvers can be used as approximations of classical systems have been explored. Operator aliasing, the quantum potential, interference, and momentum aliasing can produce spurious evolution. These limitations put constraints on the choices of simulation parameters. 

Developing a method that would adaptively resize or change simulation parameters may provide a more robust solver. Given the potential usefulness of representing $2n$ degrees of freedom in an $n$ dimensional wave function, it would be interesting to produce solvers that more carefully adhere to classical solutions. More work comparing the accuracy of the von Neumann solver to full phase space solvers was not included in this work but would further explore the utility of this extension. 

It is apparent that field solvers can in certain limits reproduce classical solutions. All of our methods approximate the classical solutions in some regimes at some resolutions. It is not immediately obvious which systems are most efficient for which regimes, a more detailed analysis from an information content perspective would be useful.

\begin{acknowledgments}
This work was performed in part under DOE Awards No. DE-AC02-76SF00515. M.K. was supported by the European Research Council under the European Unions Seventh Framework Programme (FP7/2007-2013) / ERC Grant No. 617656 ``Theories
and Models of the Dark Sector: Dark Matter, Dark Energy and Gravity’' (the primary investigator is Constantinos Skordis). 
\end{acknowledgments}

\appendix*

\section{Units}
\label{sec:units}

In this Appendix, we discuss how we create a set of dimensionless units for length, velocity, and time, which define our code units. All positions and lengths are measured in units of the simulation box, $L$, such that $0\leq x_s\leq 1$. For certain test problems, we also adopt $-0.5\leq x_s\leq 0.5$. All perturbation length scales $\lambda$ for the plasma problems are defined in these length units. Next, we cast Poisson's equation as 
\begin{equation}
    \nabla^2 V = C\rho\,,
\end{equation}
 noting that $C$ can be greater or less than $0$. $C>0$ represents attractive forces, while $C<0$ represents repulsive forces. In the electromagnetic systems, we define a frequency in the following manner:
\begin{align}
    \omega_0 = \sqrt{\frac{-C}{2}} \, .
\end{align}
We also define a time and velocity unit as 
\begin{align}
    t_0 &= \frac{2 \pi}{\omega_0} \, , \\
    u_0 &= \frac{\lambda \omega_0}{\sqrt{2} \pi} \, .
\end{align}
Note that the unit for velocity is motivated by the critical velocity in two stream instability. We normalize the total mass in the system to unity, i.e. $M_{\rm tot} \equiv 1$. For the plasma problems, we can also define a unit of electric field.
\begin{align}
    E_0 = \frac{C}{M_{\rm tot}} \, .
\end{align}
Using these definitions we can define the following dimensionless simulation units for length, mass, time, velocity, and electric field respectively
\begin{align}
    &x_s = \frac{x}{L} \in [0,1] \textit{ or } [-.5,.5] \,,\\
    &m_s = \frac{m}{M_{\rm tot}} \in [0,1] \,,\\
    &t_s = \frac{t\sqrt{-C}}{2\pi \sqrt{2}} = \frac{t}{t_0} \,,\\
    &u_s = \frac{2\pi u}{\lambda \sqrt{-C}} = \frac{u}{u_0} \,.
\end{align}
For our nonclassical solvers, we also need to choose a value for $\Tilde{\hbar} \equiv \hbar/m_{\rm part}$ which describes the relationship between position and momentum in these solvers.

\section{von Neumann generalization to 2D} \label{sec:3D}

In this Appendix, we discuss how to generalize the von Neumann solver to higher dimensions by example with the two-dimensional case. The fundamental object, the density matrix, $P$, is a $2n$ dimensional object still given by the sum of the outer product of stream wave functions, which can be written element wise as

\begin{equation}
    P_{iljm} \equiv \frac{1}{N_s} \sum^{N_s}_{s} \psi_{ij,s} \psi^*_{lm,s}
\end{equation}

$\tilde{P}$ can then be calculated using a series of Fourier transforms. Here we use Einstein summation notation to describe discrete Fourier transforms on a given axis i.e.

\begin{align}
    \tilde{\rho}_{\tilde{i}} &= F_{\tilde{i}i}\rho_i \equiv \sum_{i} \Delta x \, e^{-ik_{\tilde{i}}x_i} \rho_i \, , \\
    \tilde{\rho}_{\tilde{i}} &= F^{-1}_{\tilde{i}i}\rho_i \equiv \sum_{i} \Delta x \, e^{ik_{\tilde{i}}x_i} \rho_i \, .
\end{align}

Using this notation, we write

\begin{equation}
    \tilde{P}_{\tilde{i}\tilde{l}\tilde{j}\tilde{m}} \equiv F^{-1}_{\tilde{i}i}F^{-1}_{\tilde{j}j} F_{\tilde{l}l} F_{\tilde{m}m}  P_{iljm}
\end{equation}

The density is given using the generalization of the diagonal

\begin{equation}
    \rho_{ij} = P_{iijj} \, ,
\end{equation}

we can write the element wise position and momentum updates as 

\begin{align}
    \Tilde{P}_{iljm,t+\Delta t} &= \Tilde{P}_{iljm,t} \exp \left[  \frac{i \, \Delta t}{2 \, m \, \hbar} (p_{ij}^2 - p_{lm}^2) \right] \nonumber \\
    P_{iljm,t+\Delta t} &=  P_{iljm,t} \exp \left[  \frac{i \, \Delta t}{\hbar} (V_{lm} - V_{ij}) \right]
\end{align}

\bibliography{BIB}

\end{document}